\documentclass[a4paper,11pt]{article}
\usepackage{pos}
\title{Clustering and visualization tools to study high dimensional parameter spaces: B anomalies example}
\ShortTitle{High-dimensional visualization}

\author[a]{Ursula Laa}
\author*[b]{German Valencia}

\affiliation[a]{University of Natural Resources and Life Sciences,  Department of Landscape, Spatial and Infrastructure Sciences, Institute of Statistics,\\ Peter-Jordan-Straße 82/I, 1190 Vienna, Austria}

\affiliation[b]{School of Physics and Astronomy, Monash University,\\
  Wellington Road, Clayton, VIC-3800, Australia}

\emailAdd{ursula.laa@boku.ac.at}
\emailAdd{german.valencia@monash.edu}

\abstract{We describe the applications of clustering and visualization tools using the so-called neutral B anomalies as an example. Clustering permits parameter space partitioning into regions that can be separated with some given measurements. It provides a visualization of the collective dependence of all the observables on the parameters of the problem. These methods highlight the relative importance of different observables, and the effect of correlations, and help to understand tensions in global fits. The tools we describe also permit a visual inspection of high dimensional observable and parameter spaces through both linear projections and slicing.}

\FullConference{%
  Corfu Summer Institute 2022 "School and Workshops on Elementary Particle Physics and Gravity",\\
  28 August - 1 October 2022\\
  Corfu, Greece}

\begin{document}
\maketitle

\section{Introduction}

Many problems in physics contain large numbers of parameters and/or large numbers of predictions that are hard to visualize. Here we discuss how tours can assist with the visualization of these problems. Another issue that arises in multi-parameter problems is that of mapping different parameter regions to different prediction regions. To address this question we propose a partitioning of parameter space based on clustering predictions in observable space. To be specific we discuss the application of these tools to the so-called "neutral B-anomalies" problem, illustrating what can be learned beyond the usual global fits \cite{Laa:2021dlg}. 

The results from the tour methods that we use are usually presented as movies or animations which are not visible in the pdf file. Some of the animations that we mention here can be generated by running the example in the Shiny app https://github.com/uschiLaa/pandemonium. For the remainder, you can contact one of us directly. Short movies showing the animations referenced here can also be obtained from the arXiv version of this document.

\begin{figure}[h]
\centering{
\includegraphics[scale=0.4]{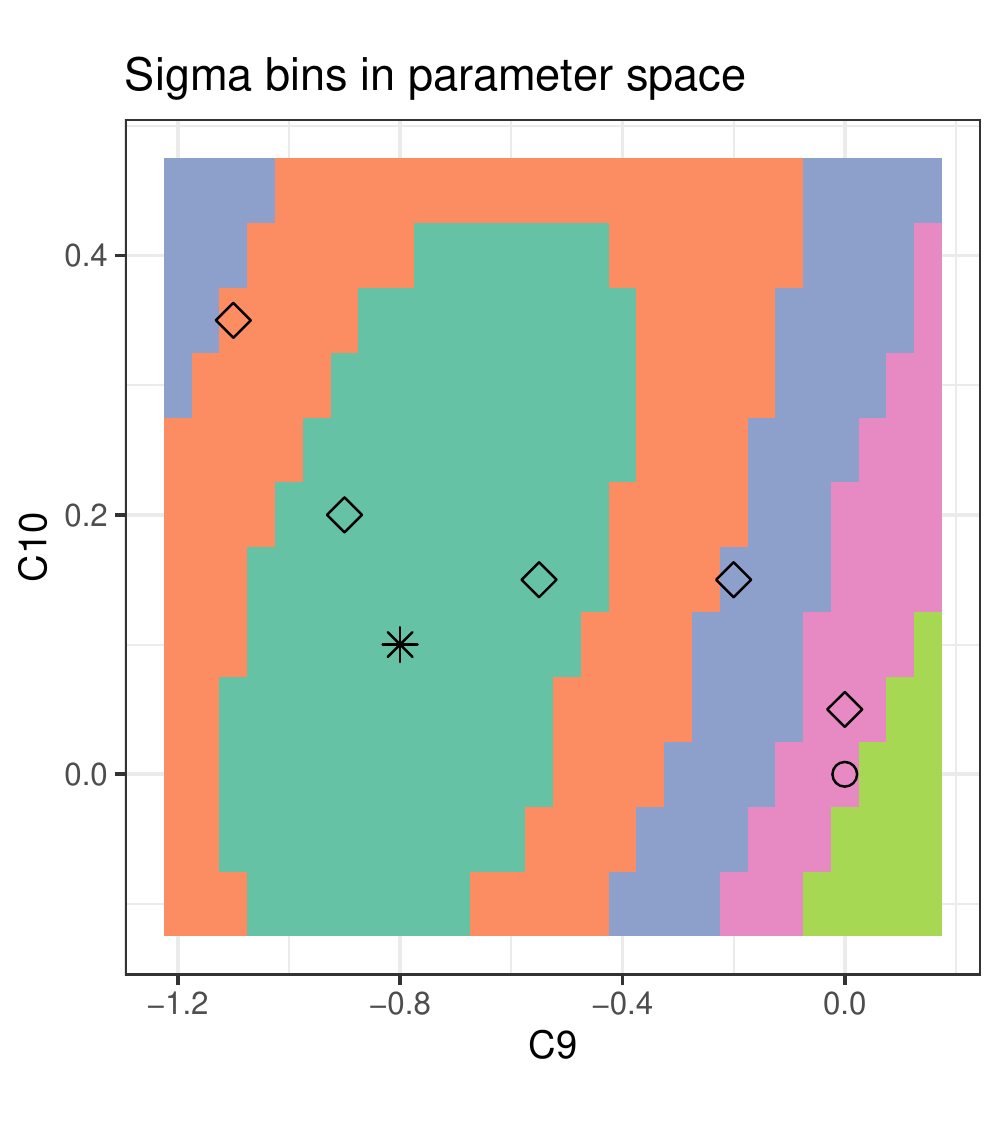}\includegraphics[scale=0.3]{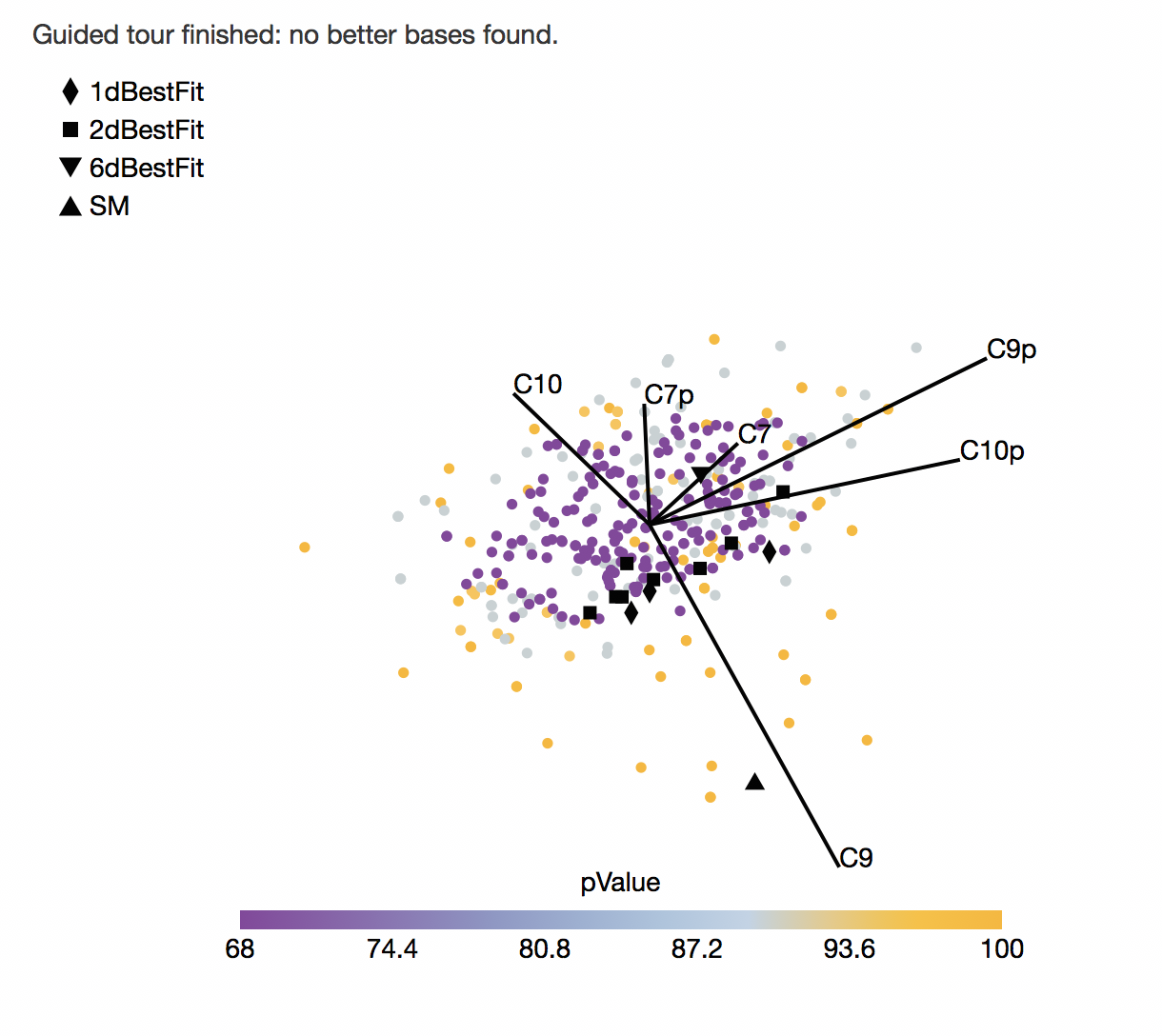}}
\caption{Left panel: partitioning of parameter space using confidence level regions from a global fit. Right panel: optimal projection of the 6d parameter space of a global fit obtained with a guided tour showing that the best fit deviates from the SM mostly along the $C_9$ direction.}
\label{f:tourGF}
\end{figure}

As we know, there are multiple observables (several hundred binned branching ratios and decay distributions) in B-meson decay modes originating from the quark level transition $b\to s\ell^+\ell^-$ where the leptons are muons or electrons. These have received a considerable amount of attention due to persistent deviations from the standard model (SM), although recently the discrepancies in two of the observables ($R_K$ and $R_{K^*}$) seem to have disappeared \cite{LHCb:2022qnv}. 
This system has been studied using global fits of the hundreds of observables in terms of between two and six parameters. The results that one can obtain from that type of exercise include
\begin{itemize}
\item finding the best-fit (BF) parameters
\item measuring the goodness of the fit and comparing it to the SM
\item model selection to find the subset of parameters that can best describe the data
\item finding confidence level intervals for the fitted parameters
\end{itemize}
The latter already corresponds to a partitioning of parameter space based on a single distance to a reference point (the experimental values of the observables), as illustrated in the left panel of Figure~\ref{f:tourGF}. These results are very useful for physics studies to determine whether a given model is a suitable description of the data. Even for these existing studies, a visualization of the high dimensional confidence level regions can provide information beyond what is observed by considering two-dimensional projections. As an example, we show in the right panel of Figure~\ref{f:tourGF}, the result of a guided tour used to find the projection illustrating the largest separation between the SM point and the best fit to the data from a six-parameter fit from 2019 \cite{Capdevila:2018jhy}. This view indicates that the apparent deviation from the SM occurs along the $C_9$ direction in parameter space. Even more  intuition can be gained from animation~1, which shows a grand tour of the 6D region near the BF and we have marked the SM, the 6D BF point and several one and two-dimensional fits described in \cite{Capdevila:2018jhy}.

\section{Beyond global fits}

Some of the new insights into a data set that can be obtained from clustering are
\begin{itemize}
\item A partitioning of parameter space into clusters uses all inter-point distances. It does not depend on a specific reference point, such as an experimental measurement that may not yet exist (or that may change, as was recently the case with $R_K$). Different clustering parameters are suitable to emphasize different aspects of the problem. 
\item The number of clusters, or different groups, in the space, reflects the resolving power of a specific data set.
\item The clustering results can help isolate trends and effects from subsets of observables.
\end{itemize}
In addition, high-dimensional visualization tools can offer new perspectives. For example, they
\begin{itemize}
\item Permit a visual inspection of the collective dependence of the observables on the parameters.
\item Provide a graphic display of observable spaces with more than three dimensions.
\item Highlight the relative importance of different observables which can help prioritize further studies.
\item Provide a virtual assessment of the impact of correlations, dominant observables, tensions in global fits, and others.
\end{itemize}

\subsection{The B-anomaly example}

For conceptual clarity and to simplify the visualization, we first select a subset of the observables and parameters that have been used in the literature to discuss the $b\to s\ell^+\ell^-$ system. Most existing global fits treat the Wilson coefficients (WC) in an effective Hamiltonian as free parameters. We will first illustrate our methods with a two-dimensional case where $C_9^\mu$ and $C_{10}^\mu$ are the parameters, later on, we add two more parameters $C_{9^\prime}^\mu$ and $C_{10^\prime}^\mu$ for a four-dimensional example. The effective weak Hamiltonian responsible for the $b\to s\ell^+\ell^-$ transitions at the B-mass scale is usually written as
\begin{align}
{\cal H}_{\rm eff} =& -\frac{4G_F}{\sqrt{2}}V_{tb}V^\star_{ts}\sum_{i}C_{i}^\ell(\mu){\cal O}_{i\ell}(\mu)\\
{\cal O}_{9} ^\ell=& \frac{e^2}{16\pi^2}(\bar{s}\gamma_{\mu}P_Lb)(\bar\ell \gamma^\mu\ell),~{\cal O}_{9^\prime}^\ell = \frac{e^2}{16\pi^2}(\bar{s}\gamma_{\mu}P_Rb)(\bar\ell \gamma^\mu\ell), \\
{\cal O}_{10}^\ell =& \frac{e^2}{16\pi^2}(\bar{s}\gamma_{\mu}P_Lb)(\bar\ell \gamma^\mu\gamma_5\ell),~{\cal O}_{10^\prime}^\ell = \frac{e^2}{16\pi^2}(\bar{s}\gamma_{\mu}P_Rb)(\bar\ell \gamma^\mu\gamma_5\ell).
\end{align}
where we have singled out the four operators we discuss here. This set of operators, with real WC, only allows CP-conserving new physics and affects  only the muons. Our notation is such that these WC refer exclusively to new physics, they are 0 in the SM, and the SM effects are accounted for separately. 

The dimensionality of observable space also needs to be reduced for clarity. We select a subset of fourteen observables based on the ranking analysis of \cite{Capdevila:2018jhy}. 
These observables are listed in Table~\ref{t:obs}, where the last column gives the ID that this observable had in \cite{Capdevila:2018jhy}. We note, however, that some definitions of the observables are not identical: the sign of $P_2$ is reversed here, and in some cases, different experimental measurements are being averaged as we rely on {\tt flavio} \cite{Straub:2018kue} for this study. We choose the observables marked with a $\color{red}{\star}$ which were singled out as the most important ones for the determination of $C_9^\mu$ and $C_{10}^\mu$ in the global fits. We also include the ones marked with a $\color{blue}{\star}$ which were singled out as important for $C_{9^\prime}^\mu$ and $C_{10^\prime}^\mu$. The remainder $P_2$ and $P_5^\prime$ bins are chosen to complete the $q^2$ distributions for these two observables. Note that $R_K$ and $R_{K^\star}$ are the ones whose experimental values have recently changed and this will provide us with a chance to evaluate this change within this study. The experimental values are taken from:
for $P_5^\prime$ LHCb \cite{Aaij:2020nrf}, CMS \cite{Sirunyan:2017dhj} and ATLAS \cite{Aaboud:2018krd}; $P_2$ LHCb \cite{Aaij:2020nrf}; $R_K$  LHCb \cite{Aaij:2019wad} and Belle \cite{Abdesselam:2019lab}; $R_{K^\star}$ LHCb \cite{Aaij:2017vbb} and Belle \cite{Abdesselam:2019wac}. The corrected values of $R_K$ and $R_{K^\star}$ \cite{LHCb:2022qnv}. Unless specifically stated otherwise, all plots and results will use the "old" values of $R_K$ and $R_{K^\star}$.
\begin{table}[htp]
{ \centering
\begin{tabular}{|c|c|c|c|}  \hline
ID & Observable & Exp.  & ID in \cite{Capdevila:2018jhy} \\ \hline
1$\color{blue}{\star}$ & $P_5^\prime(B \to K^* \mu\mu) [0.1-0.98] $ &$0.52\pm 0.10$  &20 \\
2 & $P_5^\prime(B \to K^* \mu\mu) [1.1-2.5] $ & $0.36\pm 0.12$  & 28\\
3 & $P_5^\prime(B \to K^* \mu\mu) [2.5-4] $ & $-0.15\pm0.14$   & 36 \\
4 $\color{red}{\star}$ & $P_5^\prime(B \to K^* \mu\mu) [4-6] $ &$-0.39\pm0.11$ & 44 \\
5 $\color{red}{\star}$& $P_5^\prime(B \to K^* \mu\mu) [6-8] $ & $-0.58\pm0.09$  & 52 \\
6 & $P_5^\prime(B \to K^* \mu\mu) [15-19] $ & $-0.67\pm0.06$ & 60\\
7 & $P_2(B \to K^* \mu\mu) [0.1-0.98] $ & $0\pm 0.04$ &  17\\
8 & $P_2(B \to K^* \mu\mu) [1.1-2.5] $ & $-0.44\pm0.10$  & 25\\
9  & $P_2(B \to K^* \mu\mu) [2.5-4] $ & $-0.19\pm0.12$  & 33\\
10$\color{red}{\star}$ & $P_2(B \to K^* \mu\mu) [4-6] $ &$0.10\pm0.07$ &  41\\
11 $\color{red}{\star}$& $P_2(B \to K^* \mu\mu) [6-8] $ & $0.21\pm0.05$ &  49\\
12 $\color{red}{\star}$ & $P_2(B \to K^* \mu\mu) [15-19] $ & $0.36\pm 0.02$ & 57 \\
13 $\color{red}{\star}\color{blue}{\star}$& $R_K(B^+  \to K^+ ) [1.1-6] $ & $0.86 \pm 0.06$  & 98 \\
& new value &$0.949^{+0.047}_{-0.046}$ &\\
14 $\color{red}{\star}\color{blue}{\star}$& $R_{K^*} (B^0  \to K^{0 *}) [1.1-6] $ & $0.73\pm 0.11$  &  100\\
& new value &$1.027^{+0.077}_{-0.073}$ &\\
 \hline
\end{tabular} 
\caption{List of observables used to cluster measurements with an underlying $b\to s\ell^+\ell^-$ quark transition. }
\label{t:obs}
}
\end{table}
The 2D BF to this dataset as obtained from {\tt flavio} \cite{Straub:2018kue} is the point $(C_9^\mu,C_{10}^\mu)=(-0.8,0.1)$, and lies 3.7$\sigma$ from the SM. These two points are marked with an $\ast$ and an $\circ$ in most of the plots. The BF after the change in $R_K$ and $R_{K^\star}$ to this same dataset becomes $(C_9^\mu,C_{10}^\mu)=(-0.4,-0.1)$. .

For our study, we will generate models (sets of 14 predictions) on a grid of values for $(C^\mu_9,C^\mu_{10})$. The original Shiny app requires the grid to be uniform but this is not needed in general. All the predictions are generated with {\tt flavio} \cite{Straub:2018kue}  and the grid is chosen to be large enough to contain both the SM and the BF points.

\section{Clustering}

To partition the continuous parameter space we consider model points $M_k$ defined by their coordinates $(C_9^\mu,C_{10}^\mu)_k$ in parameter space and by their coordinates $(O_1,\cdots,O_{14})_k$ in observable space. It is easier (but not necessary) to use a distance function that can be calculated from coordinates. To this effect, we define the coordinates of each model point in observable space to be 
\begin{equation}
Y_{ki} = \sum_j \Sigma^{-1/2}_{ij}  (X_{kj} - R_j)\approx \sum_j \frac{1}{\sqrt{(\Sigma^{-1})_{ii}}} (\Sigma^{-1})_{ij} (X_{kj} - R_j),
\label{coordinates}
\end{equation}
where $X_{kj}$ is the prediction of model $k$ for observable $O_j$, $R_j$ is the "origin" or reference point for that observable, and $\Sigma_{ij}$ is the total covariance matrix including both theoretical and experimental uncertainties and correlations. The origin, $R_j$ is arbitrary but would typically be chosen as a special point. In this example that could be the experimentally observed point $E_i$, the SM prediction, or any other preferred model. These coordinates thus measure the distance from the reference point in units of combined theoretical and experimental uncertainty.  Using these coordinates, we define the (square of the) distance {\bf between models} as
\begin{equation}
d_{\chi^2}(X_k, X_l) = \sum_{i,j} [X_{ki} - X_{li}] (\Sigma^{exp} + \Sigma^{th})^{-1}_{ij} [X_{kj} - X_{lj}]=\sum_i(Y_{ki}-Y_{li})^2.
\end{equation}
The last equality follows if $\Sigma$ does not depend on the model, which is an often-used approximation particularly when the experimental errors dominate. In this case, the clustering results will not depend on the reference point. In particular, they {\bf would not change} as a result of the recent change in the central values of $R_K$ and $R_{K^\star}$.

\begin{figure}[h]
\centering{
\includegraphics[scale=0.18]{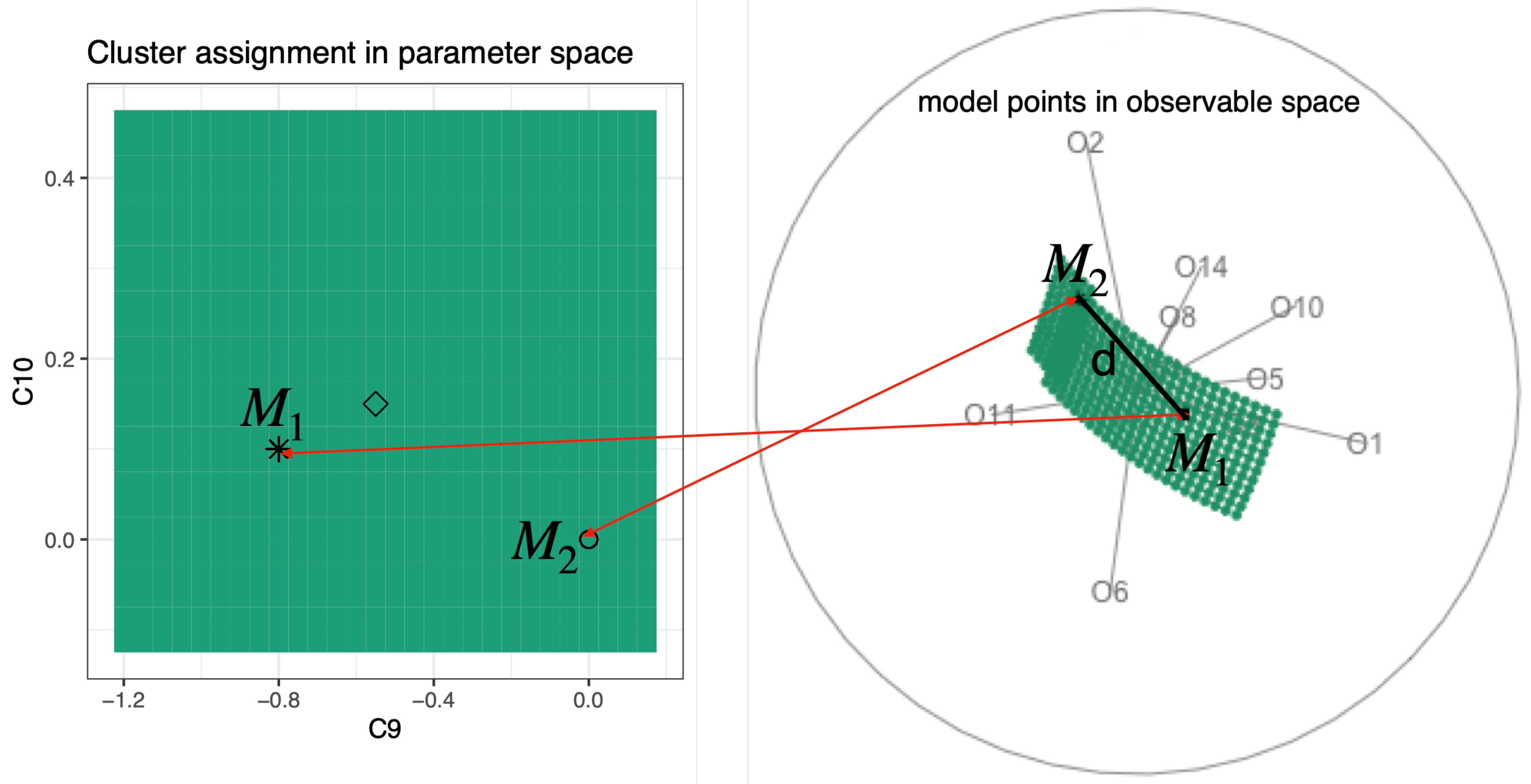}\includegraphics[scale=0.36]{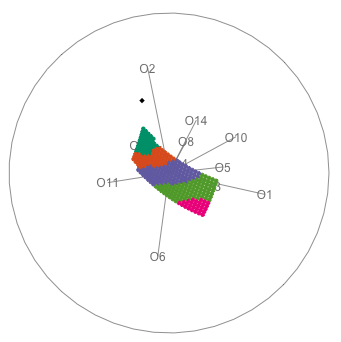}}
\caption{Partitioning the (continuous) parameter space by measuring the distance between two models $M_1$ and $M_2$ in observable space (left two panels). The result in this example is shown in the right panel and is obtained as described in the text.}
\label{f:dismap}
\end{figure}

This definition of distance is just the Euclidean distance with the coordinates defined by Eq.~\ref{coordinates}, and it can be interpreted as a $\Delta \chi^2$. We exploit this interpretation to construct the partitioning by first defining a centroid and a radius for each cluster. The centroid $c_j$ of cluster $C_j$ is the member of the cluster which minimizes $f(c,C_j)=\sum_{x_i\in C_j}d(c,x_i)^2$ and the radius of the cluster is $r_j = \max_{x_i \in C_j} d(c_j, x_i)$. The centroids are meant to be a representative point for each cluster that can serve as a benchmark for further studies. With these definitions, one can use "one-sigma" clusters, for example, to obtain the partitioning. The interpretation, in this case, is that if a future BF to all experiments falls at one of the centroids, the corresponding cluster contains all the points lying in the 1$\sigma$ region, $\Delta\chi^2\leq 2.3$ for two parameters. Note however that no one centroid is singled out by a global fit: at this stage, there is no need for a fit (or even measurements) to exist. Similarly, we can require that any two centroids be separated by at least $\Delta\chi^2> 2.3$. Two caveats are important: there will always be points as close to each other as we want that, nevertheless, sit on different clusters, and the boundaries between clusters will shift if the parameter range that is being studied is changed. This clustering method is sketched in the left two panels of Fig.~\ref{f:dismap}, and the results for our example are then shown in the third panel in observable space.

\begin{figure}[h]
\centering{
\includegraphics[scale=0.38]{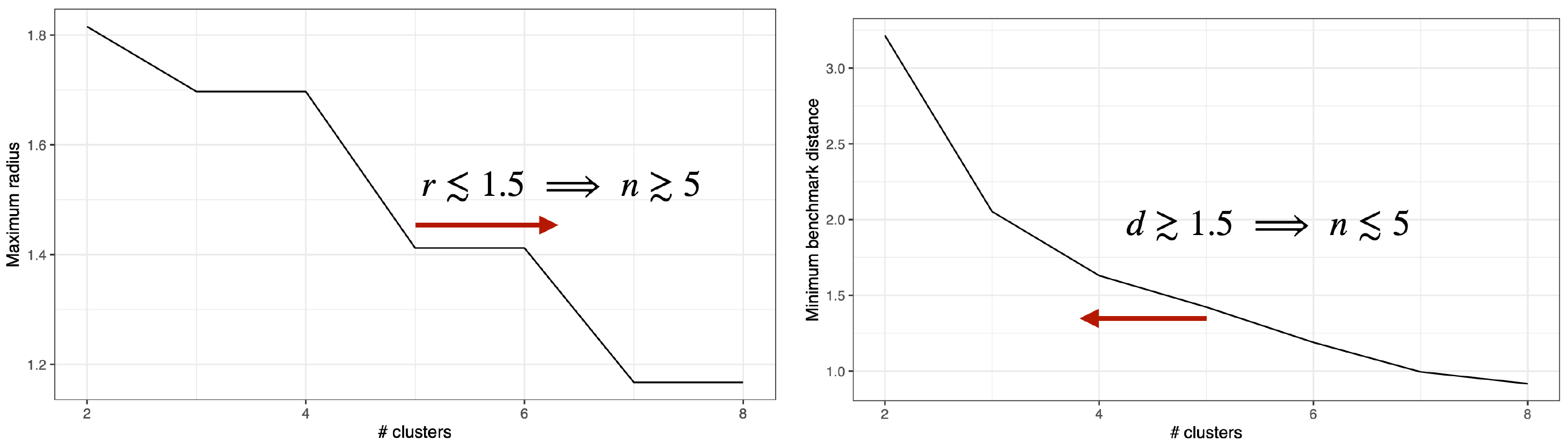}}
\caption{Maximum cluster radius and minimum distance between centroids as a function of the number of clusters determine the optimal choice for this example which is five clusters.}
\label{f:numclu}
\end{figure}

The distance between clusters is referred to as linkage, and here our focus is on Ward.D2 linkage which defines clusters by minimizing a within-cluster dissimilarity function. To decide on the number of clusters we compute both the maximum cluster radius and the minimum distance between centroids as a function of the number of clusters. The concept of a cluster as a set of points that are indistinguishable from each other at some level of confidence fixes the maximum radius and thus the minimum number of clusters. For the centroids to differ at some level of confidence, the minimal distance between them must also be fixed and this condition results in a maximum number of clusters. These combined requirements lead to there being five clusters in this example as illustrated in Fig.~\ref{f:numclu}. The resolving power of a given data set depends on the parameter space volume, the range of predictions for a given observable over that region of parameter space, and the size of the uncertainty in both measurements and predictions. It is possible to increase the resolving power by adding observables or by increasing the precision of a measurement. The latter happened with the latest measurements of $R_K$ and $R_{K^\star}$, and including this updated experimental error would improve the resolution of this set to six clusters. These changes in $R_K$ and $R_{K^\star}$ have minimal effect on the results of our clustering exercise so we proceed with the results as obtained in \cite{Laa:2021dlg}. We later show what changes occur when the new values of $R_K$ and $R_{K^\star}$ are used.

\begin{figure}[h]
\centering{
\includegraphics[scale=0.4]{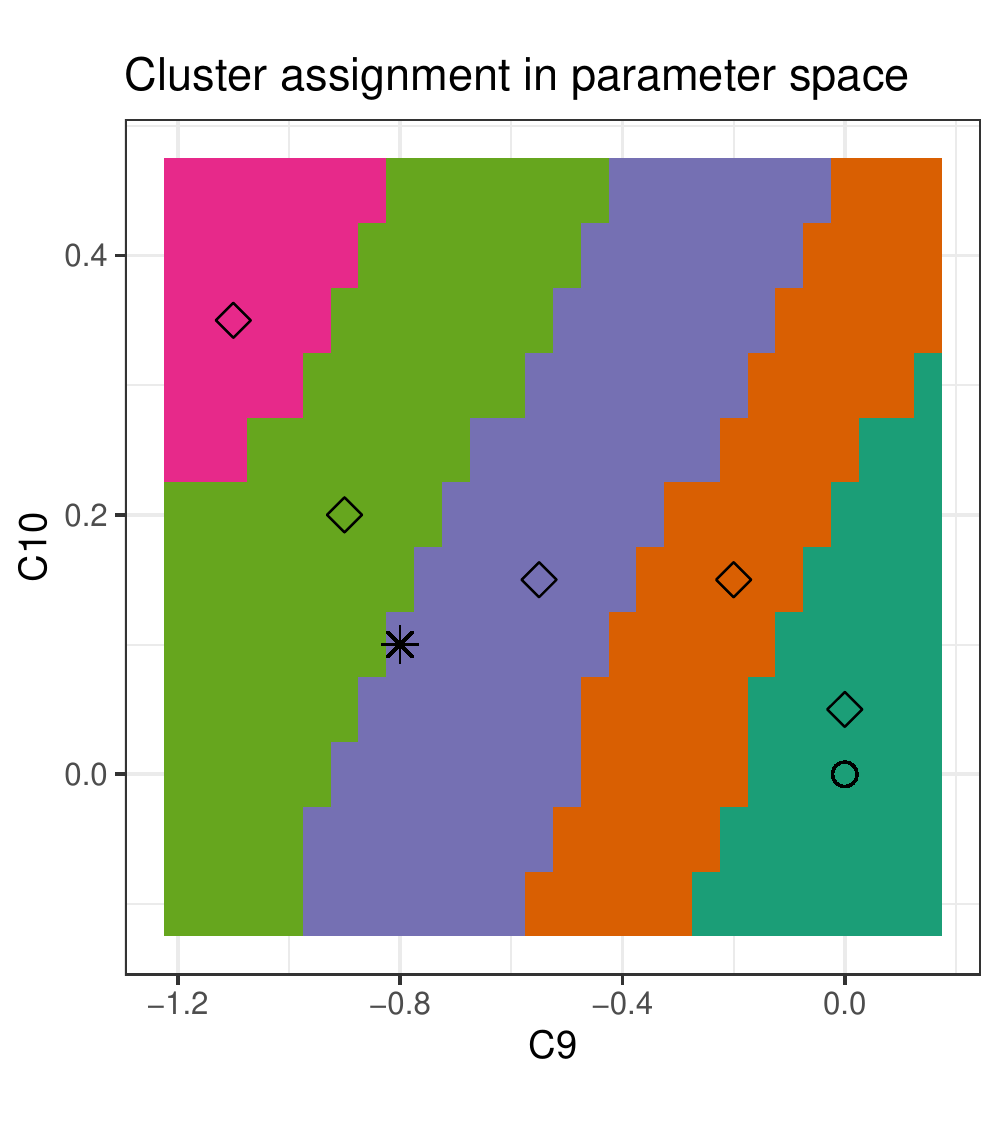}\includegraphics[scale=0.55]{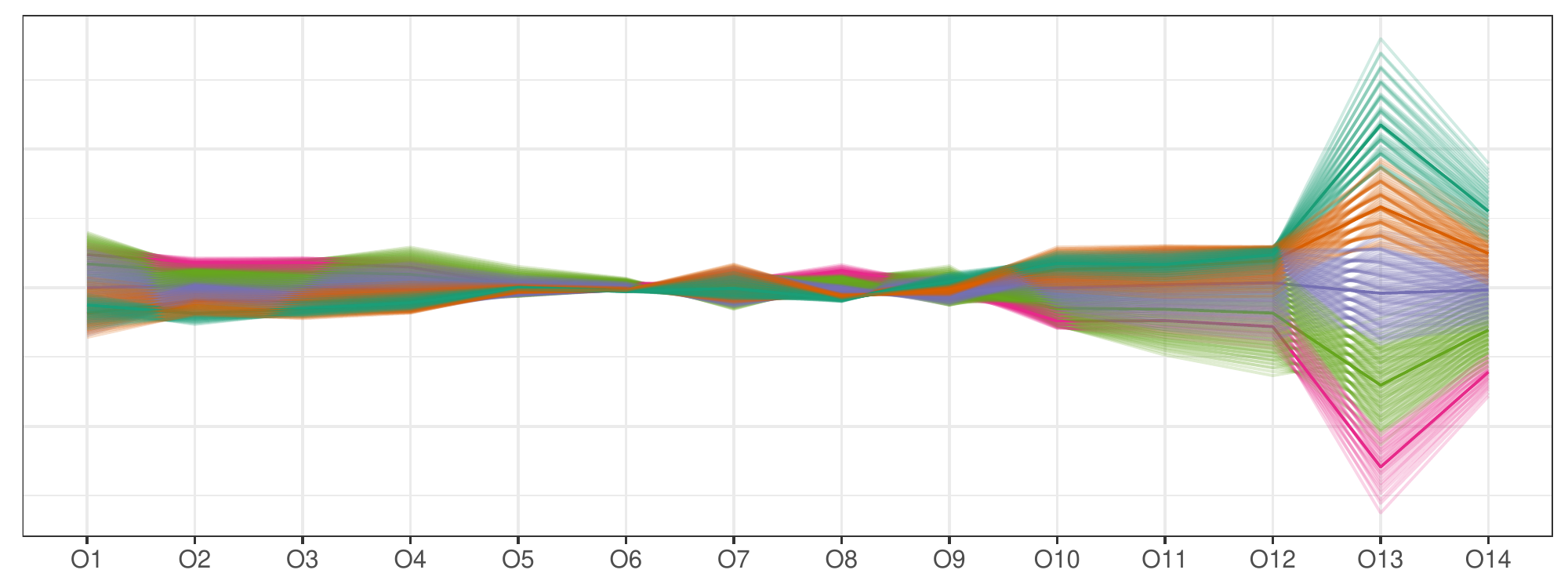}}
\caption{Clustering result using Ward.D2 linkage (which minimizes the variance within clusters) and Euclidean distance (left panel), and the corresponding centered  parallel coordinates (PC) for all 14 observables (right panel) with matching color codes. The darker line for each color in the PC plot marks the cluster benchmark (also indicated on the left, with an open diamond symbol). A projection of the 14d observable space is shown in the last panel of Fig.~\ref{f:dismap}.}
\label{f:clusres}
\end{figure}

The resulting clusters are shown in Fig.~\ref{f:clusres}. The left panel shows the partition of parameter space. The boundaries between clusters fall approximately along lines of constant $R_K$ and the significance of this will be discussed below. The right panel is a parallel coordinate (PC) plot representation of the observable space. This PC plot has been rendered after centering the coordinates. Doing this removes the information about distance from the reference point, but allows a better comparison of the relative size of variations in the predictions for each observable. If one is more interested in following the models across the plot than in the relative size of the variations, a PC plot that is centered and scaled can be used. This option also exists in the tool {\tt pandemonium}. A grand tour view of the clusters in observable space along with the experimental point (black dot) can be seen in animation 2. From that animation one can see, for example, that the experimental point is separated from  the hyperplane of predictions for all values of the two parameters.

\begin{figure}[h]
\centering{
\includegraphics[scale=0.22]{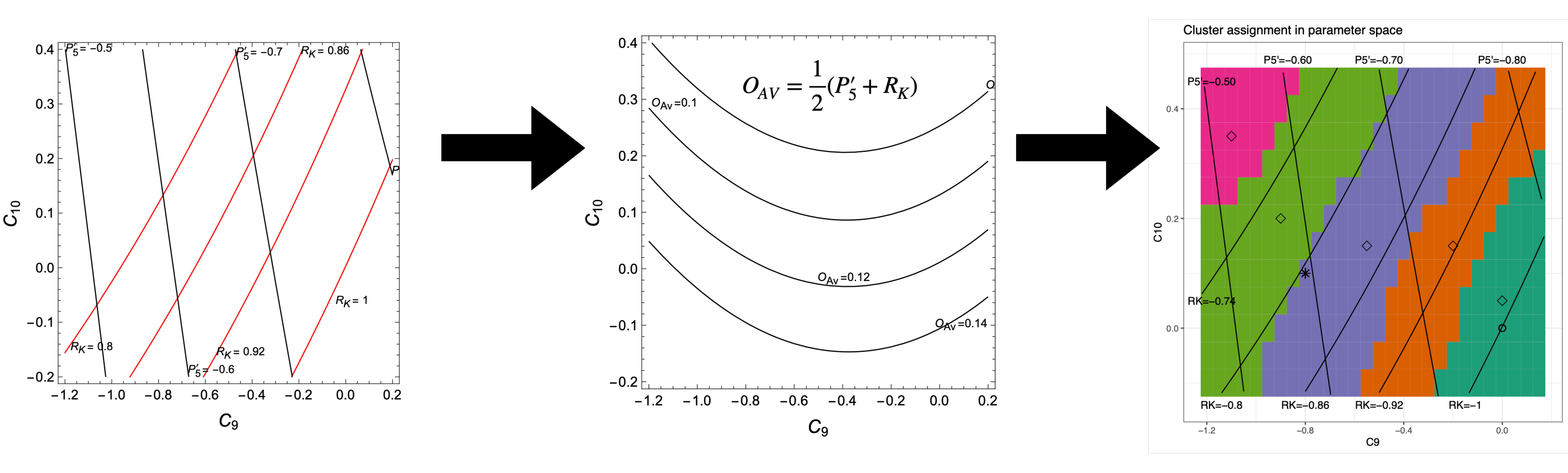}}
\caption{The left panel shows how two observables vary across the parameter region, the center panel how an average of these two varies, and the right panel the collective behavior of all 14 observables captured by the clustering result.} 
\label{f:collective}
\end{figure}

In Fig.~\ref{f:collective} we illustrate how the result of the clustering exercise helps visualize the collective dependence of all observables on the parameters. In the left panel, we show the dependence of two observables, $R_K$ (red lines mark constant values), and one bin of $P_5^\prime[4-6]$ (black lines). When there are many observables a plot like that is not very useful, instead one may want to look at combinations of observables with different weights, as illustrated in the central panel where we show the lines with constant averages of the two. The clustering exercise shown on the right panel effectively combines all the observables with different weights that can be altered by choosing a distance function and linkage. We have superimposed on this last panel the lines of constant $R_K$ and $P_5^\prime[4-6]$ to show how the boundaries between clusters follow lines of approximately constant $R_K$. This simply reflects that this observable is completely dominant in this case. This can also be seen in the PC plot of Fig.~\ref{f:clusres}. The large spread seen in $O_{13}$ in that plot reflects that, in units of uncertainty, this observable varies the most across this region of parameter space. One can also see in the same plot that $R_K$ is dominant in determining the separate clusters (almost no overlap between the colors along the $O_{13}$ coordinate. The same reasoning shows that $O_{2},O_8,O_{14}$ are also separating the clusters cleanly.

\begin{figure}[h]
\centering{
\includegraphics[scale=0.5]{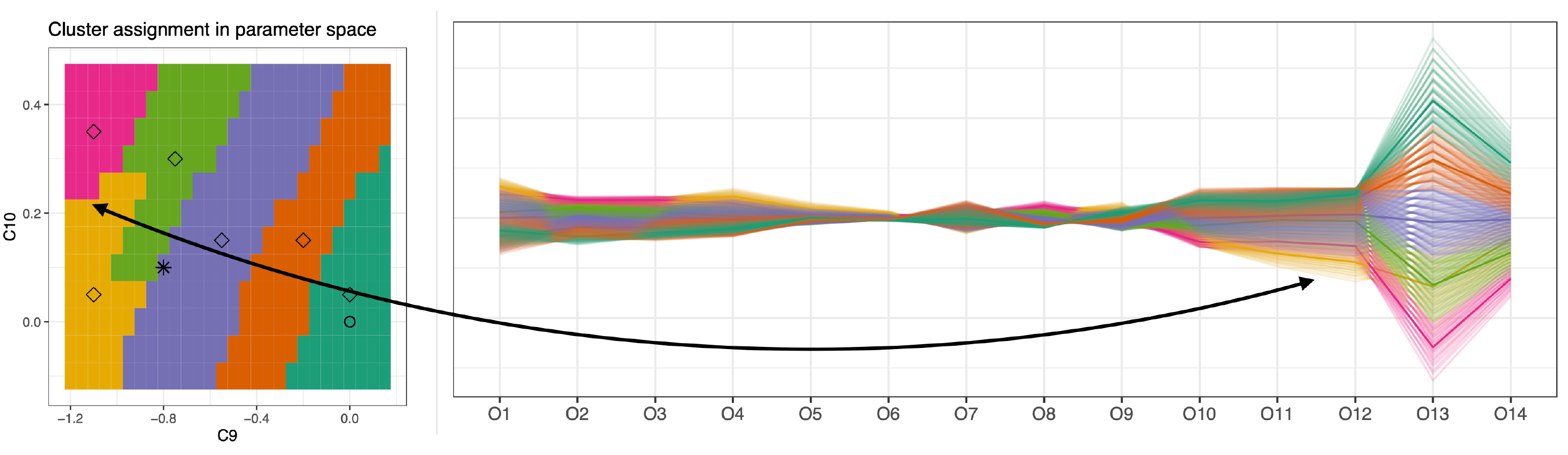}}
\caption{
When increasing the number of clusters to six we split one region, which now appears in light green and yellow. Connecting the parameter region plot (left) with the PC plot (right) we find that two observables are important for the separation of the new yellow cluster.}
\label{f:subl}
\end{figure}

Sub-leading effects can be observed by adding a sixth cluster, for example. In Fig.~\ref{f:subl} we see the sixth cluster in yellow separating from the light green by an approximately horizontal partition that indicates sensitivity to $C_{10}$ in the region away from the SM. The arrow points to the PC plot where one can see that it is mostly $O_{11,12}$ ($P_2[6-8]$ and $P_2[15-19]$) that are most important for determining the separation between yellow and pink clusters. We should caution here that numerical accuracy affects small details which at some level become just noise.

\begin{figure}[h]
\centering{
\includegraphics[scale=0.5]{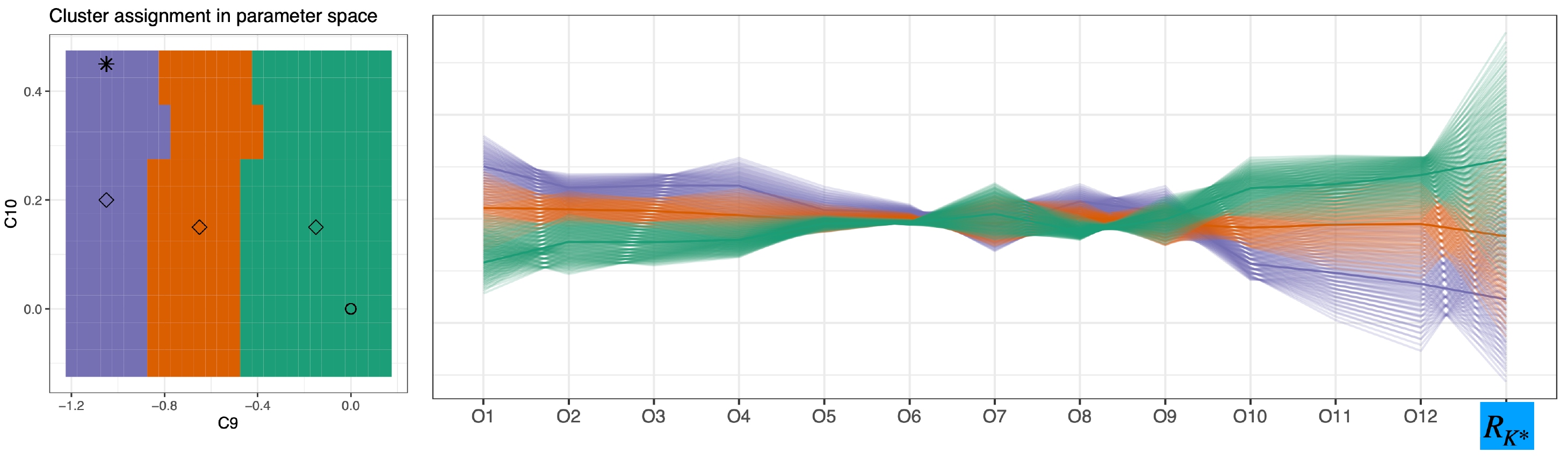}}
\caption{Clustering result after removing the dominant observable $R_{K^*}$, the thirteenth coordinate now becomes $R_{K^\star}$.}
\label{f:nork}
\end{figure}

Another way to study sub-leading effects is to remove the dominant observable, in this case, $R_K$. The result is shown in Fig.~\ref{f:nork} where we use the fact that the resolving power has been reduced to only 3 clusters. The dominant operator in the remaining set is $R_{K^*}$ but its effect is not as important as that of $R_K$. This is evident both from the size of its variation in the PC plot and from the shape of the inter-cluster boundaries. Without $R_K$, this observable set is mostly sensitive to $C_9$. The cluster separation, in this case, can be seen in the PC plot to be a collective effect due to many observables. The brown cluster is mostly due to $P_5^\prime$ and this can be seen in the PC plot which shows this cluster overlapping with others for the $P_2$ observables. Notice, of course, that the BF ($\ast$) has also shifted when we removed $R_K$.

\begin{figure}[h]
\centering{
\includegraphics[scale=0.3]{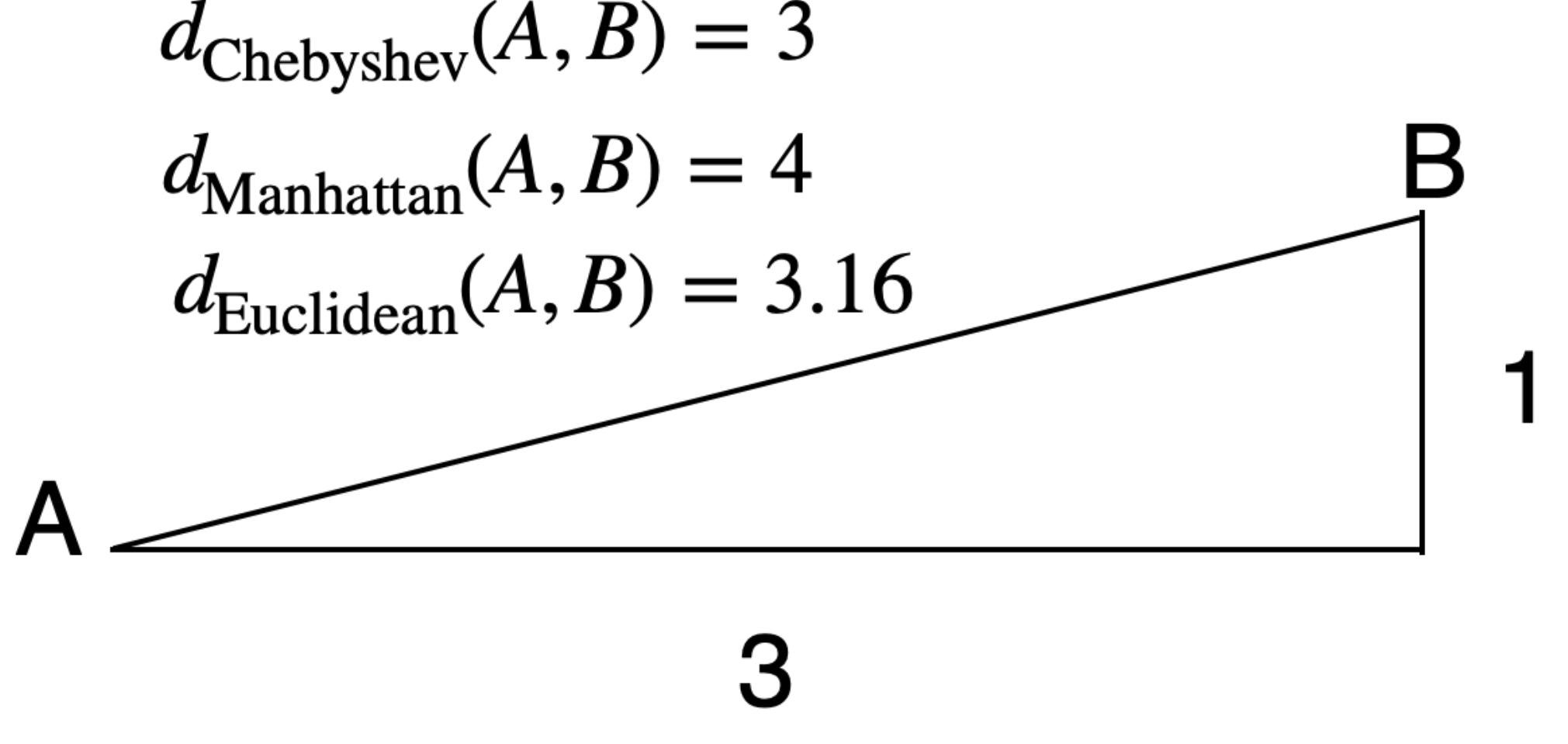}\includegraphics[scale=0.3]{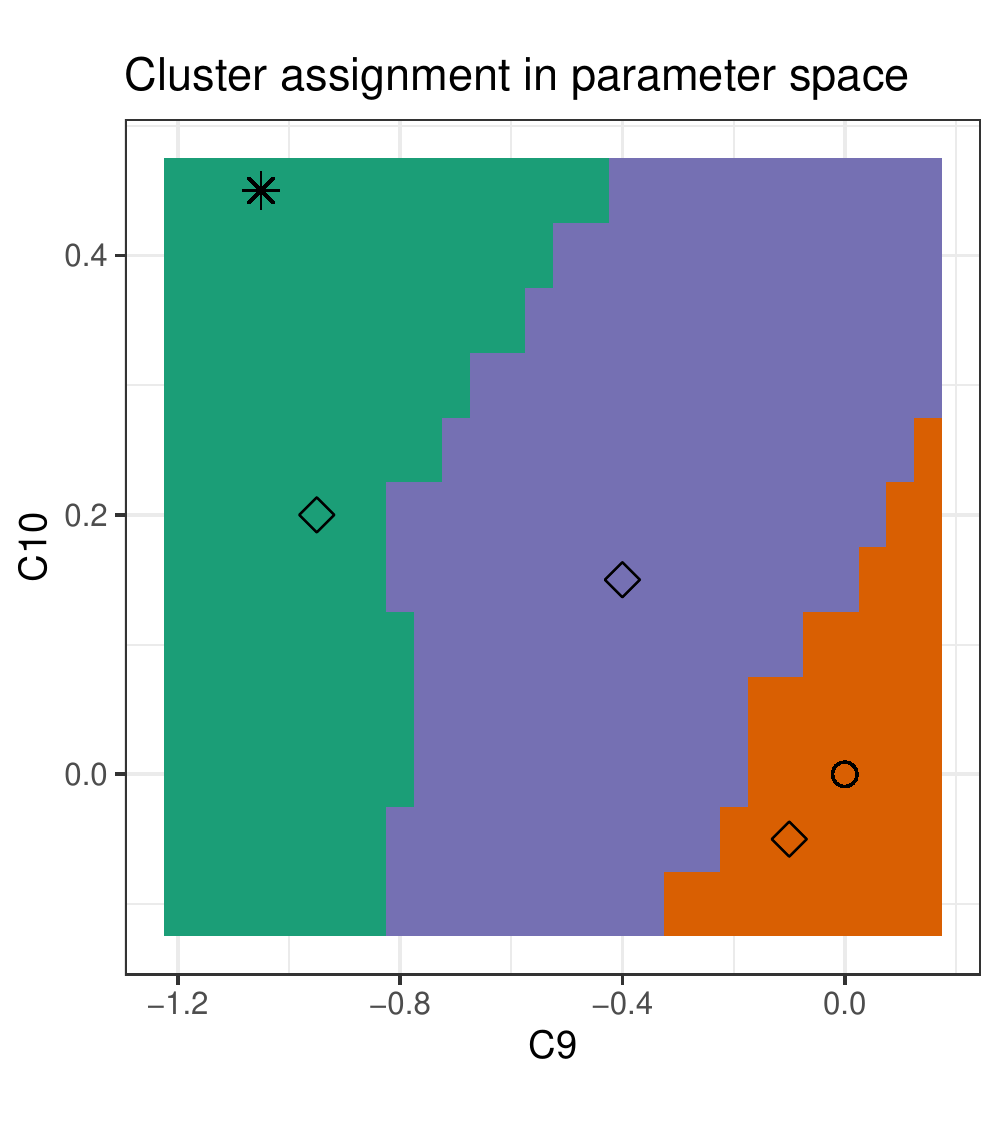}\includegraphics[scale=0.3]{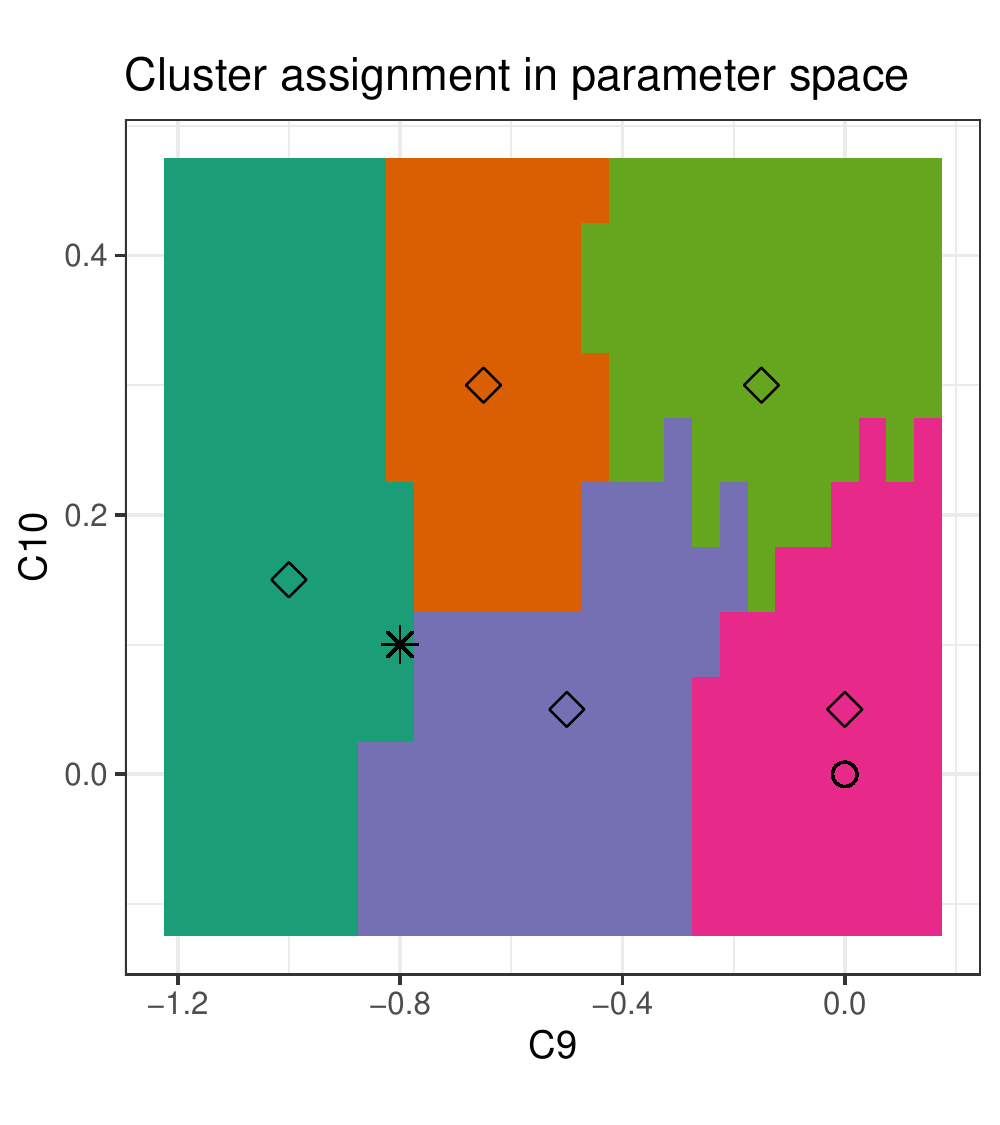}}
\caption{Left panel: sketch illustrating the difference between different distances. Center panel: observables with $R_K$ removed clustered with Chebyshev (maximum) distance and complete linkage. Right panel: all 14 observables clustered with Manhattan distance.}
\label{f:otherdistance}
\end{figure}

It is possible to enhance or suppress effects by changing the clustering parameters. To increase the importance of a dominant observable one can use maximum distance with complete linkage instead of Euclidean distance with ward linkage. The left panel of Fig.~\ref{f:otherdistance} illustrates this with a sketch in which two models, $A$ and $B$ are separated by a distance of 3 along one observable and by a distance of 1 along the other observable. Using the maximum distance removes the sub-leading observable from the picture whereas using the Manhattan distance increases its relative importance. In the center panel, we show the result of clustering our set of 13 observables (with $R_K$ removed) using maximum distance and complete linkage. This increases the weight of $R_{K^\star}$ as reflected by the change in boundary shape from that seen on the left panel of Fig.~\ref{f:nork}. The right panel is the result of clustering the full set of 14 observables but using the Manhattan distance (with Ward linkage), the clusters are now due to a collective effect.

\begin{figure}[h]
\centering{
\includegraphics[scale=0.4]{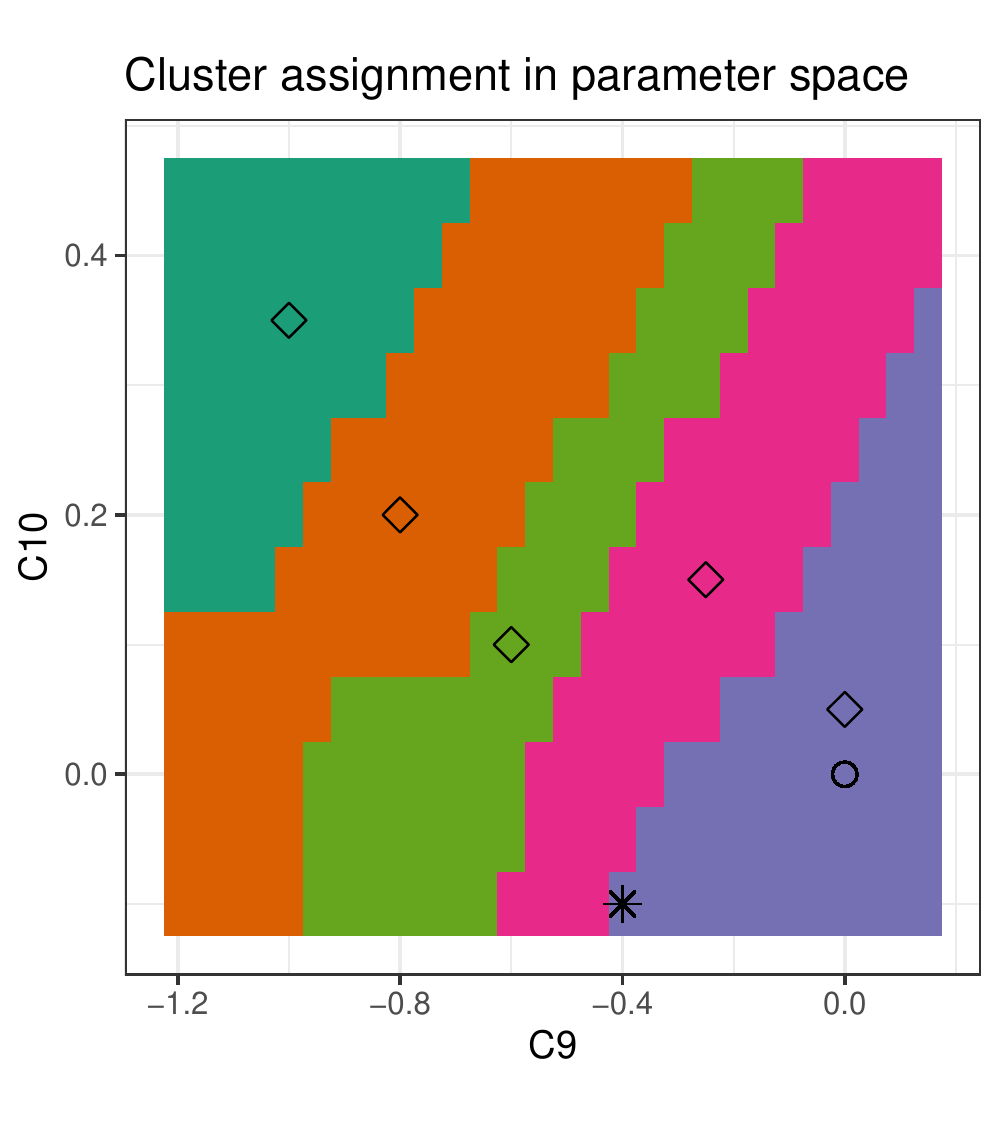}\includegraphics[scale=0.55]{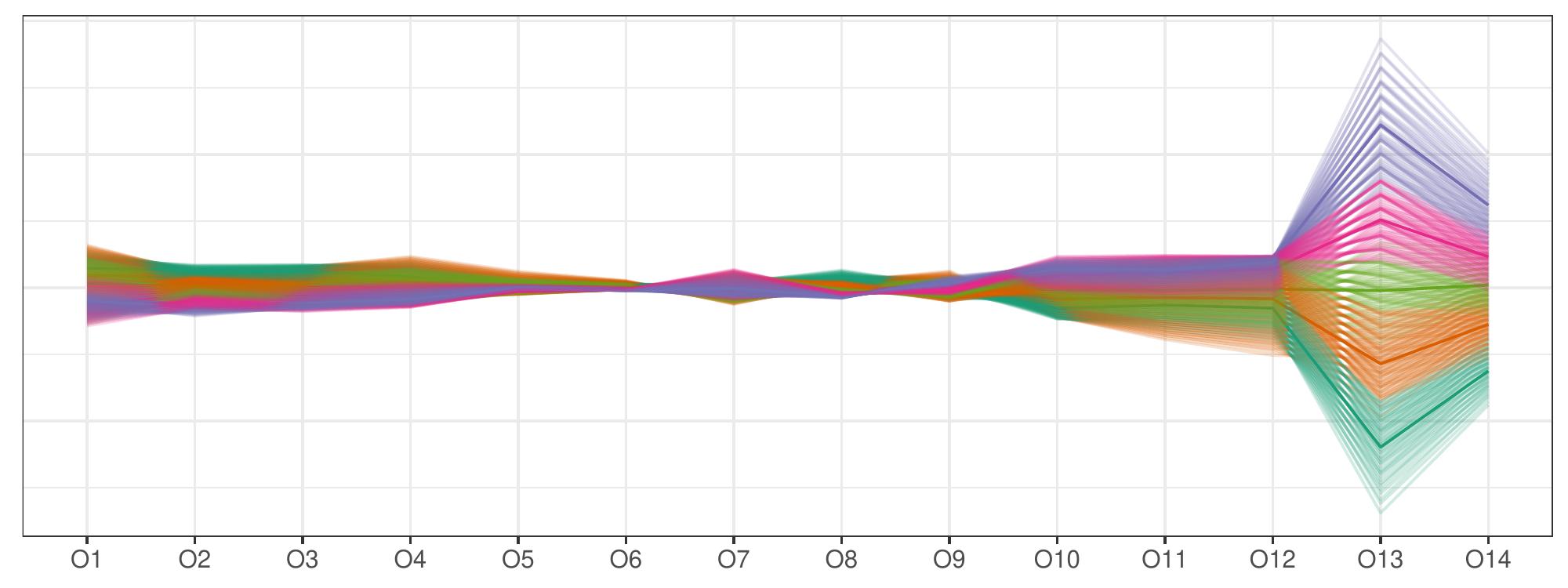}}
\caption{Clustering result matching Fig.~\ref{f:clusres} but using the new experimental values of $R_K$ and $R_{K^\star}$. }
\label{f:newrk}
\end{figure}

We end this section by using the new values of $R_K$ and $R_{K^\star}$ as recently reported by LHCb~\cite{LHCb:2022qnv}. According to our discussion, we do not expect the change in central value to alter the clustering as this does not depend on the reference point. On the other hand, the new numbers have smaller errors and this will enhance the importance of these two observables. Since they were already dominant, we do not expect any major differences. This is confirmed by comparing Fig.~\ref{f:newrk} to Fig.~\ref{f:clusres}, the shape and size of the clusters are similar but $R_K$ is even more dominant than before, the position of the BF ($\ast$) has, of course, changed.

\section{Visualization}

\begin{figure}[h]
\centering{
\includegraphics[scale=0.2]{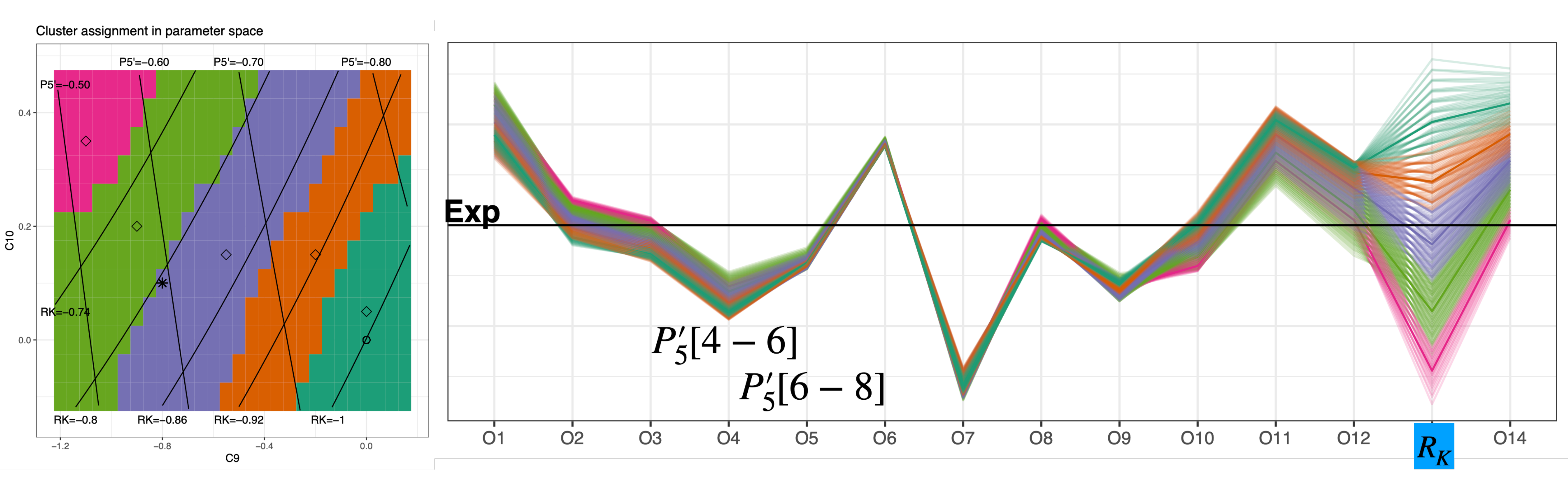}}
\caption{The left panel shows lines of constant $P_5^\prime[4-6]$ and $R_K$ superimposed on the clustering result of Fig.~\ref{f:clusres}. The right panel shows the PC plot but without centering or scaling illustrating how each observable deviates from its experimental value.}
\label{f:notcentred}
\end{figure}

The PC plots can be used to visualize other aspects of observable space if the coordinates are not centered or scaled. This is illustrated in Fig.~\ref{f:notcentred} where the horizontal line labeled "Exp" fixes the origin to the position of the experimental measurement (central value as the uncertainties are accounted for in the definition of the coordinates). This  figure allows for visual inspection of several points:
\begin{itemize}
\item We see which observables are in tension with model predictions, for example, $O_1$ cannot match the experimental value for any values of the parameters in the region of study (within some uncertainty that we quantify in the vertical axis of Fig.~\ref{f:2dvs4d}).
\item We see which observables are insensitive to the parameters $C_9$ and $C_{10}$, they are $O_6$ and $O_7$ as they exhibit minimal variation across the range studied.
\item We observe the tensions in the fit: for example, the BF lies on the boundary between purple and light green clusters. The PC plot shows that $O_4 (P_5^\prime[4-6])$ and $O_5 (P_5^\prime[6-8])$, which are the $P_5^\prime$ bins that show the largest discrepancy between the SM and experiment, prefer models within the light green cluster which have larger negative $C_9$. Recall that the experimental value of $P_5^\prime[4-6]=-0.39\pm0.11$, and thus lies outside, to the left, of the parameter region plotted. On the other hand, the pre-2022 value of $R_K$ prefers the purple cluster. One can further see that the model points that take $P_5^\prime[4-6], P_5^\prime[6-8]$ closest to their experimental value, take $R_K$ furthest away. Interestingly this tension has only become {\bf worse} with the new value of $R_K$ which agrees with the SM and would sit on the dark green cluster in this plot.
\end{itemize}

\begin{figure}[h]
\centering{
\includegraphics[scale=0.3]{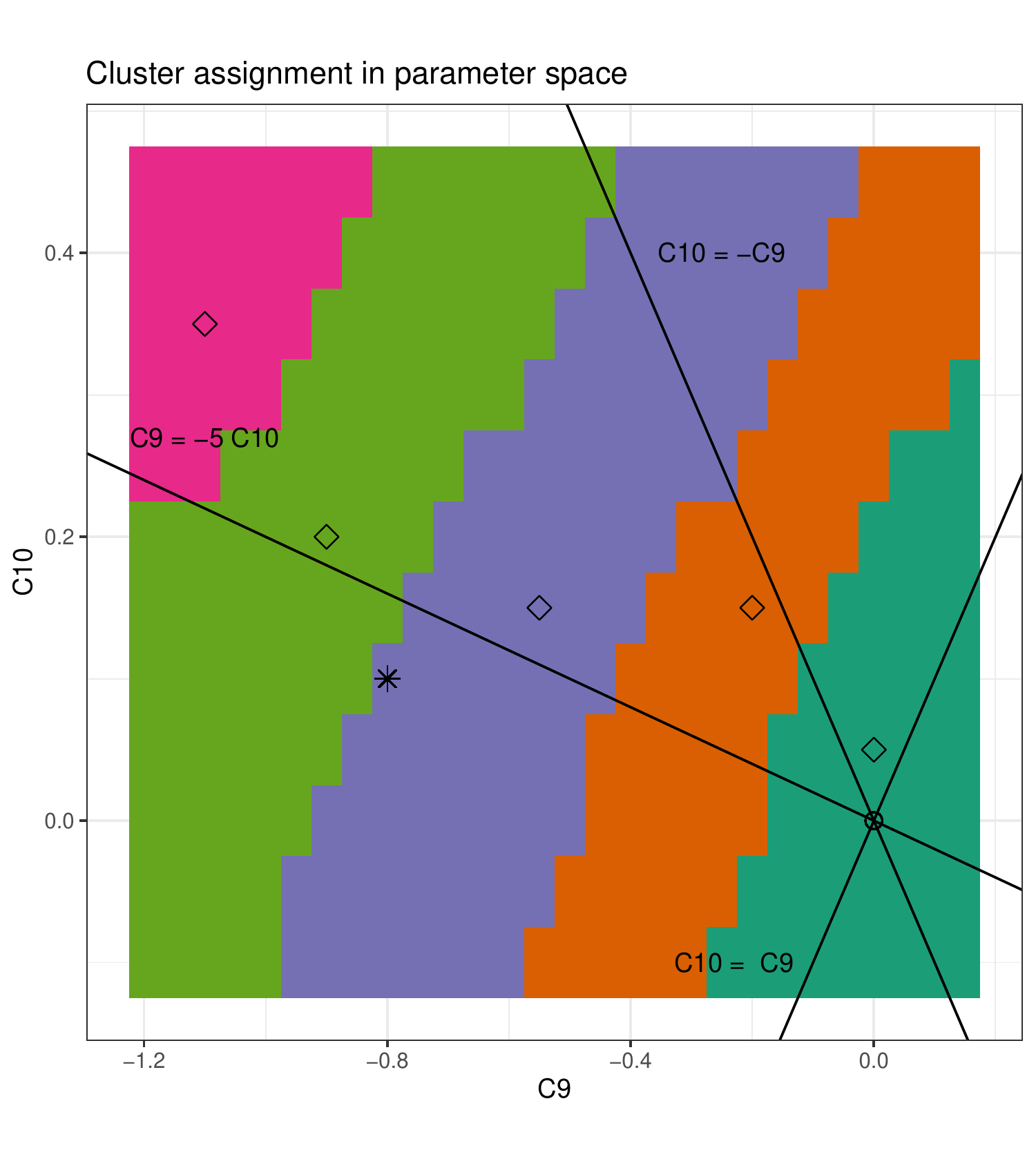}\includegraphics[scale=0.45]{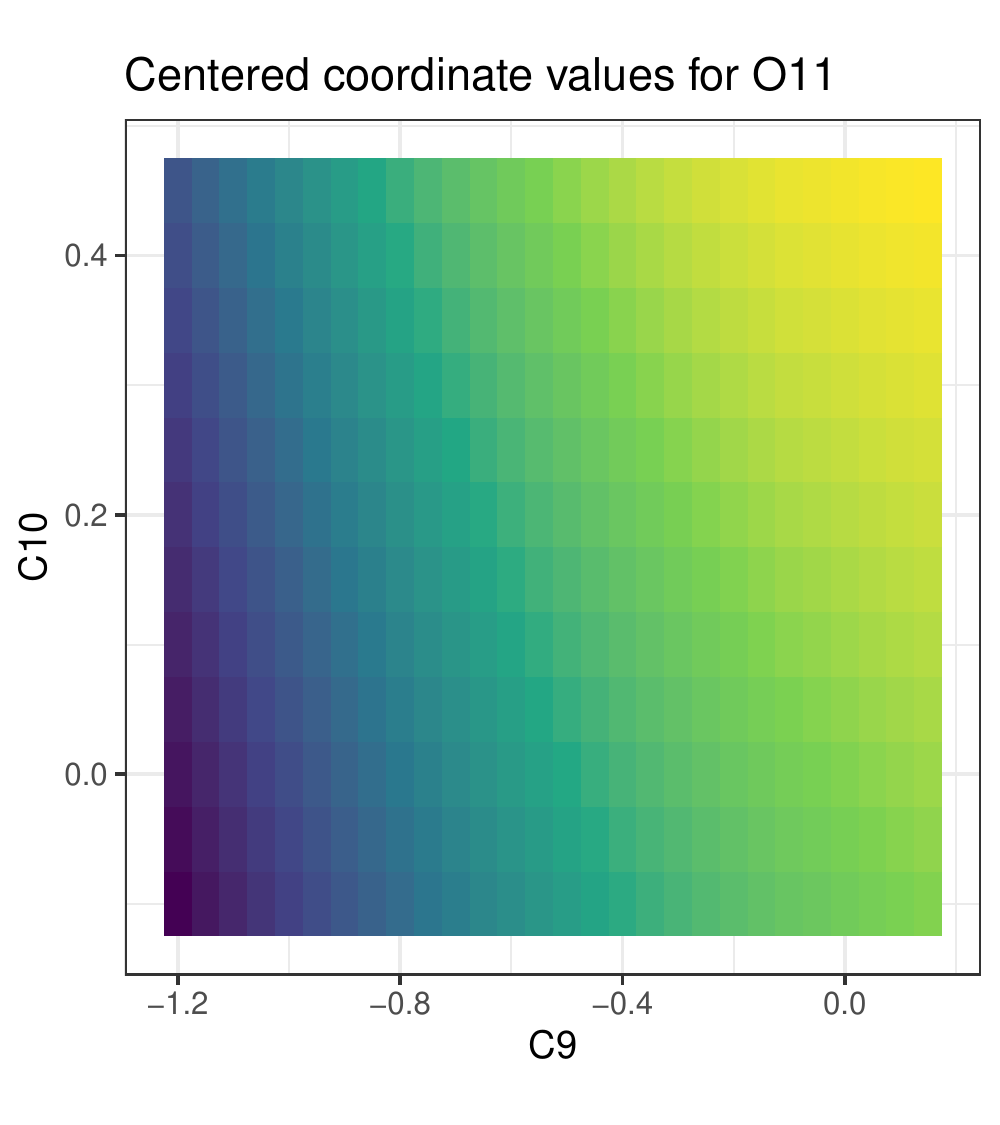}}
\caption{The left panel shows the lines $C_9=-5C_{10}$, $C_{10}=C_9$ and $C_{10}=-C_9$ superimposed on the clustering result of Fig.~\ref{f:clusres}. The right panel shows variation of $O_{11}$ ($P_2[6-8]$) across the parameter range.}
\label{f:sensitivity}
\end{figure}

The sensitivity of the observable set to given directions in parameter space can be studied and correlated with the variation of specific observables across the parameter range. For example, in Fig.~\ref{f:sensitivity}, the superimposed lines show that the set is mostly sensitive to models with $C_{10}\approx 0.2 C_9$, and that it has almost no sensitivity to models where $C_{10}=C_9$. Both of these features were already known from the results of global fits and this approach offers a clear visual picture. The right panel shows $O_{11}$ which varies across the parameter range in an orthogonal manner (this one is selected from the interactive tool {\tt pandemonium} which displays all of them), indicating that one way to increase sensitivity to models with $C_{10}=C_9$ is to improve the precision in the measurement of $O_{11}$ ($P_2[6-8]$).

\begin{figure}[h]
\centering{
\includegraphics[scale=0.2]{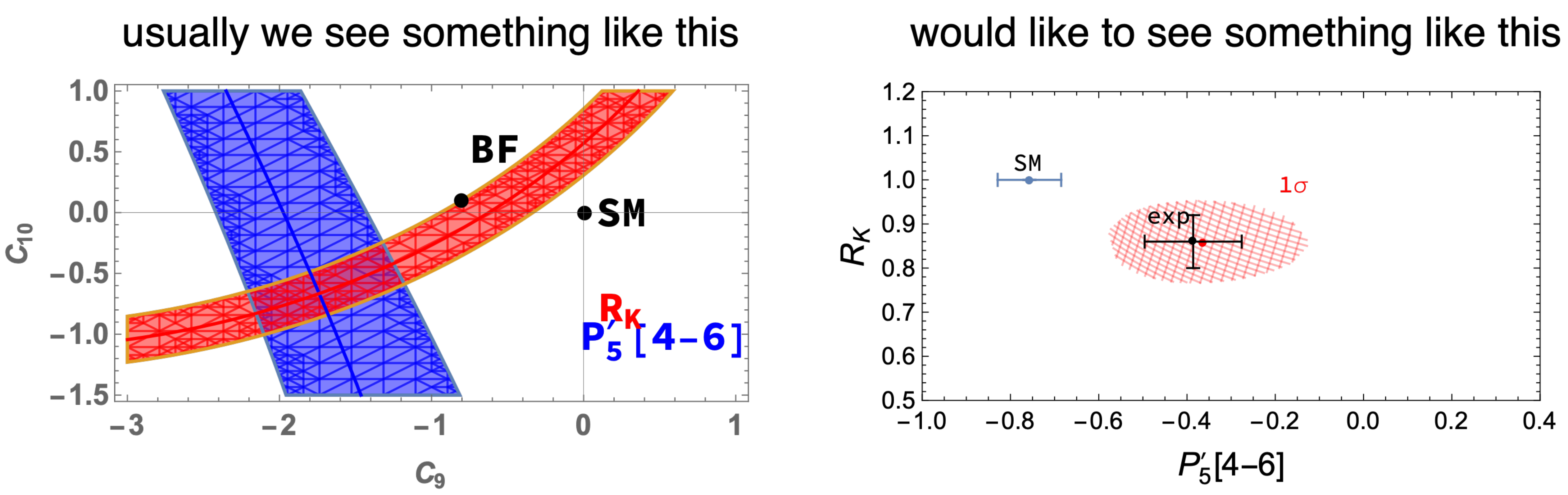}\includegraphics[scale=0.2]{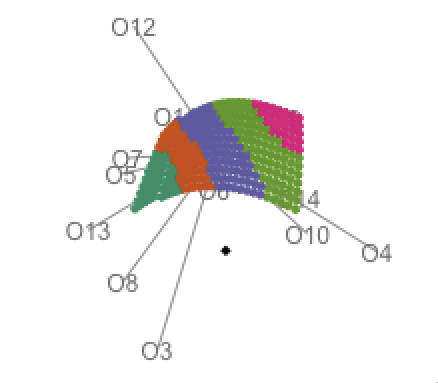}}
\caption{The left and center panels contrast the information that can be conveyed by parameter and observable space displays. The right panel is a projection of the 14-dimensional observable space partitioned into five clusters that shows clearly how the experimental point (black $\blacklozenge$) is separated from all the models parameterized by this range.}
\label{f:visual_obs}
\end{figure}

Tours allow us to visualize the high-dimensional (14 in this example) observable space and see how models compare to the measurement. On the left panel of Fig.~\ref{f:visual_obs} we illustrate a typical 2D plot in parameter space and contrast it with the corresponding 2D plot in observable space. The two convey complementary information, with the latter revealing the relative position of a model prediction and the measurements. To do this in high dimensions is possible using PC plots such as the one in the right panel of Fig.~\ref{f:notcentred}, but also using tours. Tours give a more intuitive idea of the full space as can be seen in animation~2. In the right panel of Fig.~\ref{f:visual_obs} we show one projection from the grand tour of the animation. This indicates that this parameter space cannot reach the experimental point.

\section{The case with four parameters}

\begin{figure}[h]
\centering{
\includegraphics[scale=0.1]{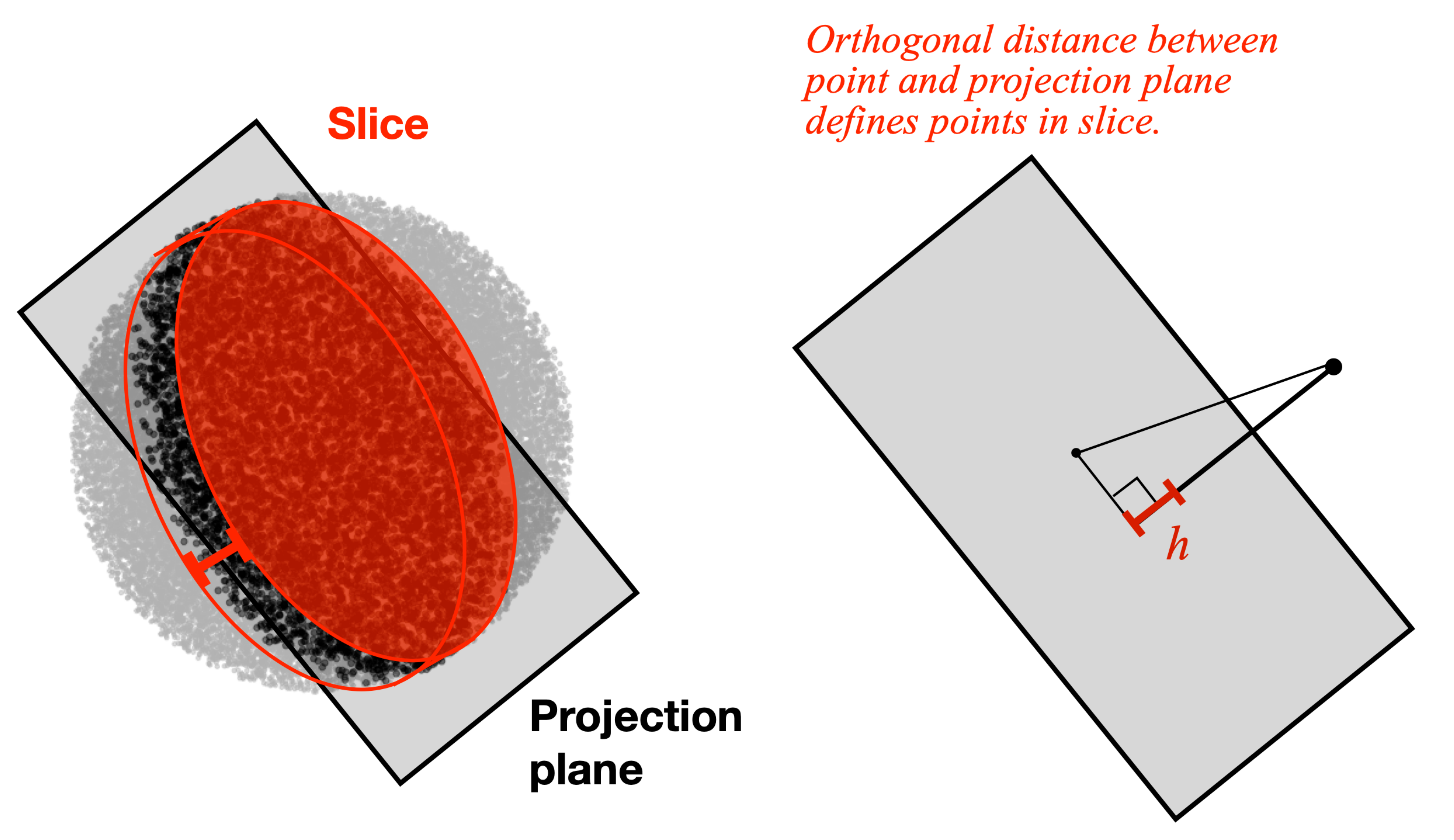}}
\caption{Sketch of how a slice of high-dimensional data can be selected based on the orthogonal point distance from the projection plane.}
\label{f:slice}
\end{figure}

From the physics perspective, including the two additional parameters $C_{9^\prime}^\mu$ and $C_{10^\prime}^\mu$, allows the exploration of models with right-handed quark currents. These are interesting in their own right but are disfavored by global fits. From the visualization perspective, the problem is complicated by the presence of two high-dimensional (more than three) spaces. This additional complication requires the introduction of slicing tools  \cite{Laa:2019bap,Laa:2020wkm}  to inspect 2D projections of thin slices in the orthogonal space, as suggested in Fig.~\ref{f:slice}.

\begin{figure}[h]
\centering{
\includegraphics[scale=0.22]{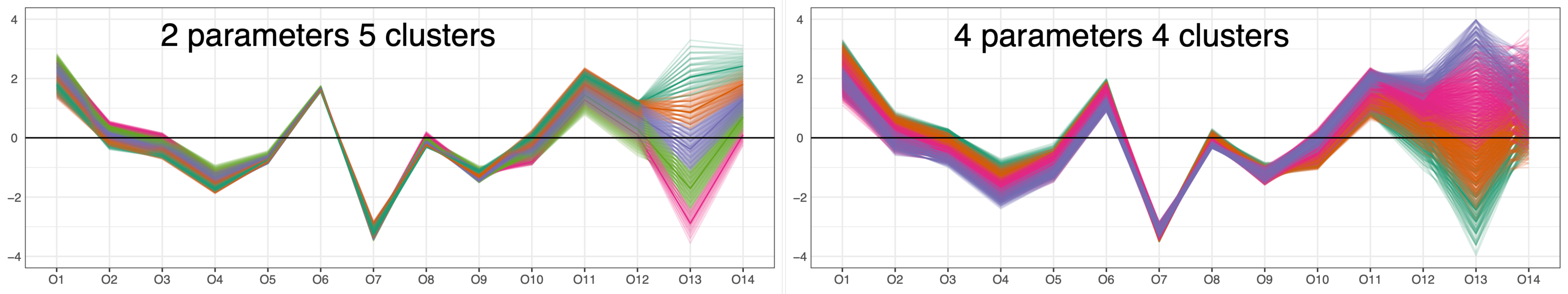}}
\caption{PC plots for two parameters and five clusters (left) and four  parameters and four clusters (right) obtained with Ward linkage and Euclidean distance. The plots are aligned to match the vertical scale.}
\label{f:2dvs4d}
\end{figure}

In our B-anomalies example, we enlarge our parameter space of study choosing ranges for the two new parameters that cover both the SM and at least their  1$\sigma$ ranges around the BF found in global fits. With the new parameter space and the same 14 observables, the resolution is only four clusters and we compare this case to the two-parameter case using PC plots for both cases in Fig.~\ref{f:2dvs4d}. We can immediately see that the extended range of predictions increases the overlap with the experiments (both plots have the same vertical scale). One can see, in particular, that the range of predictions for $O_{4,5,6}$ extends towards the origin with the enlarged parameter space. This would be evidence (within errors, of course) for $C_{9^\prime}^\mu$ and $C_{10^\prime}^\mu$ being necessary to account for the data. Looking at $O_{13}$ we see that $R_K$ no longer cleanly separates the clusters. We also observe a reduced tension between $P_5^\prime$ and $R_K$.

\begin{figure}[h]
\centering{
\includegraphics[scale=0.3]{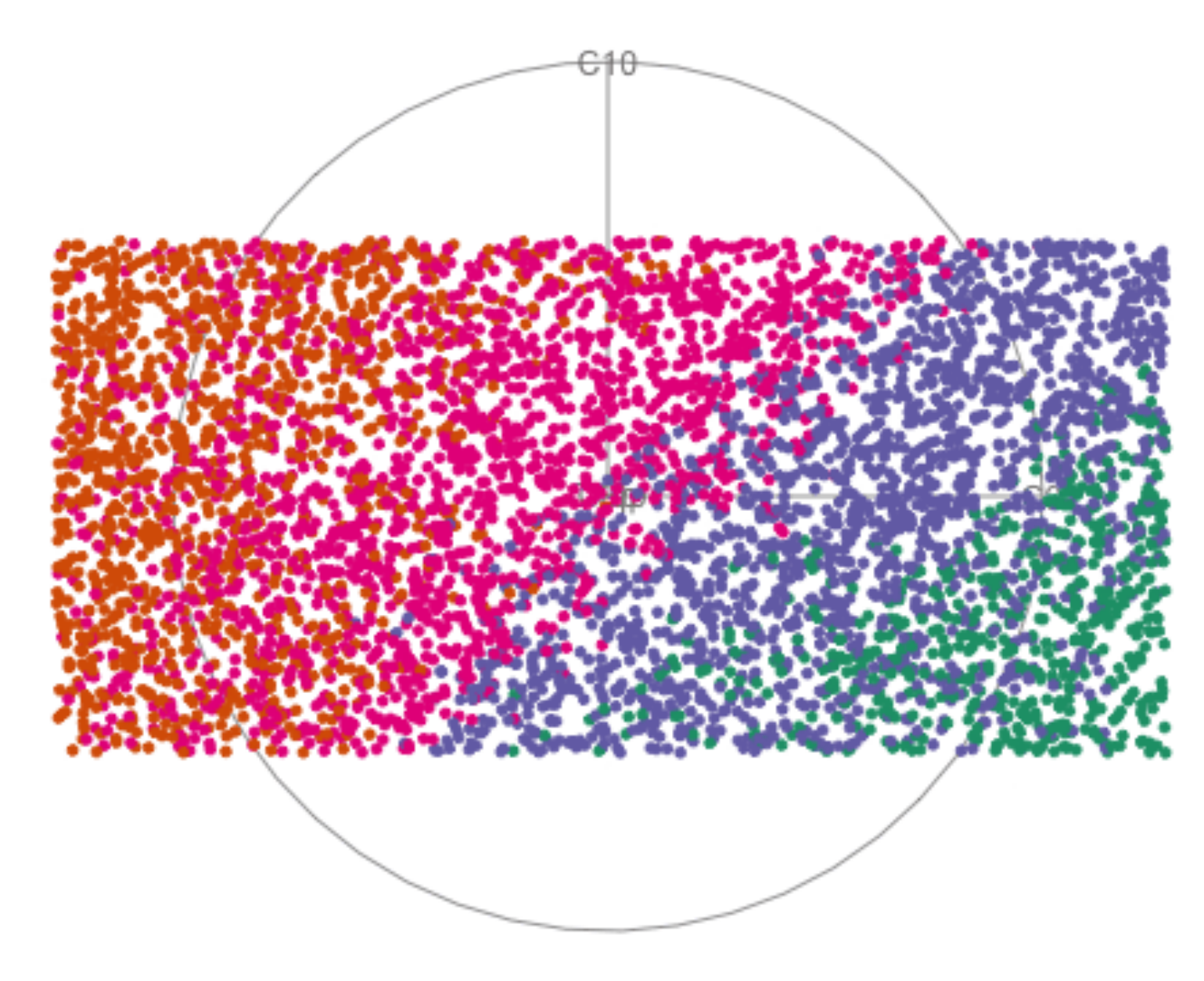}\includegraphics[scale=0.24]{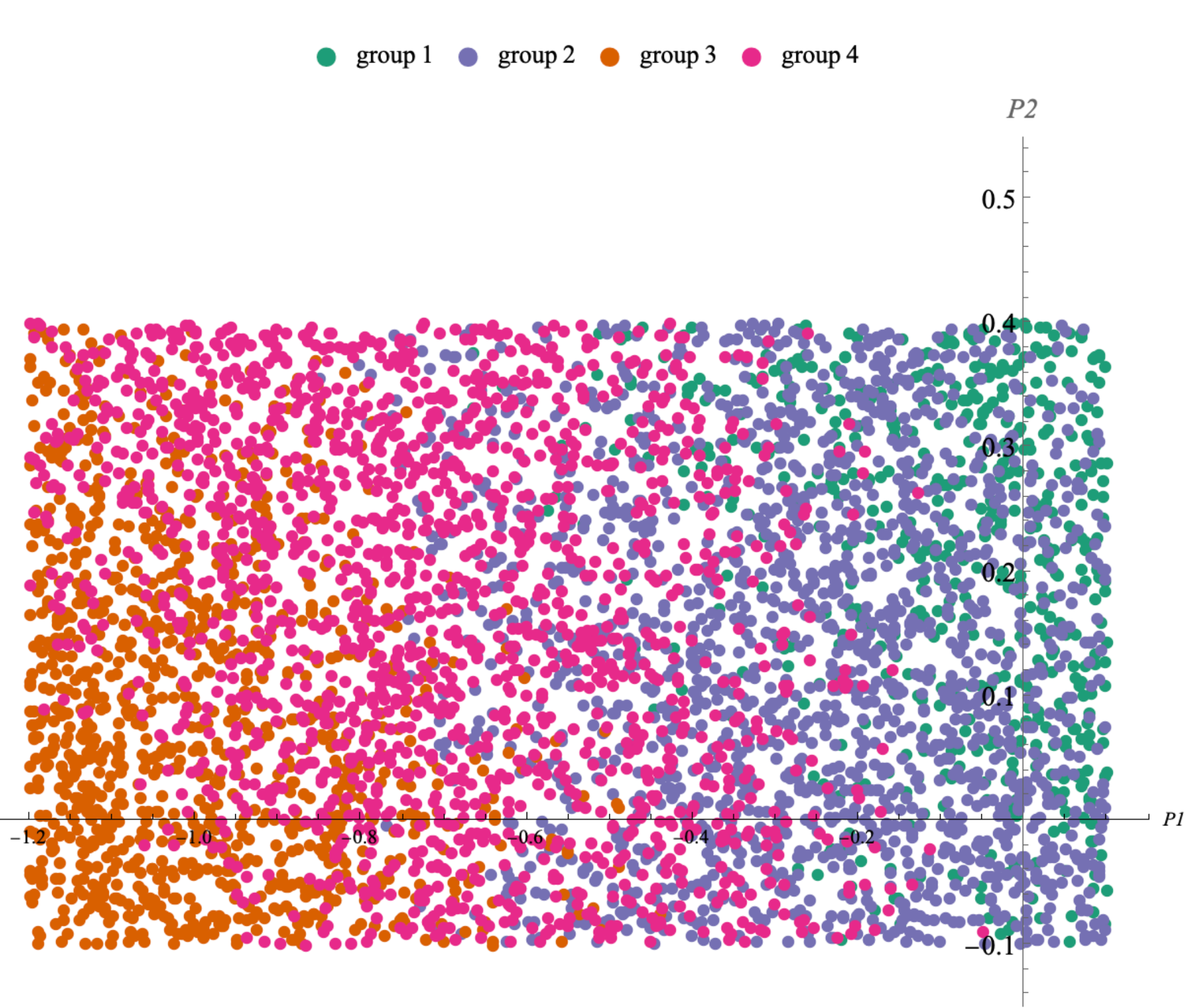}\includegraphics[scale=0.3]{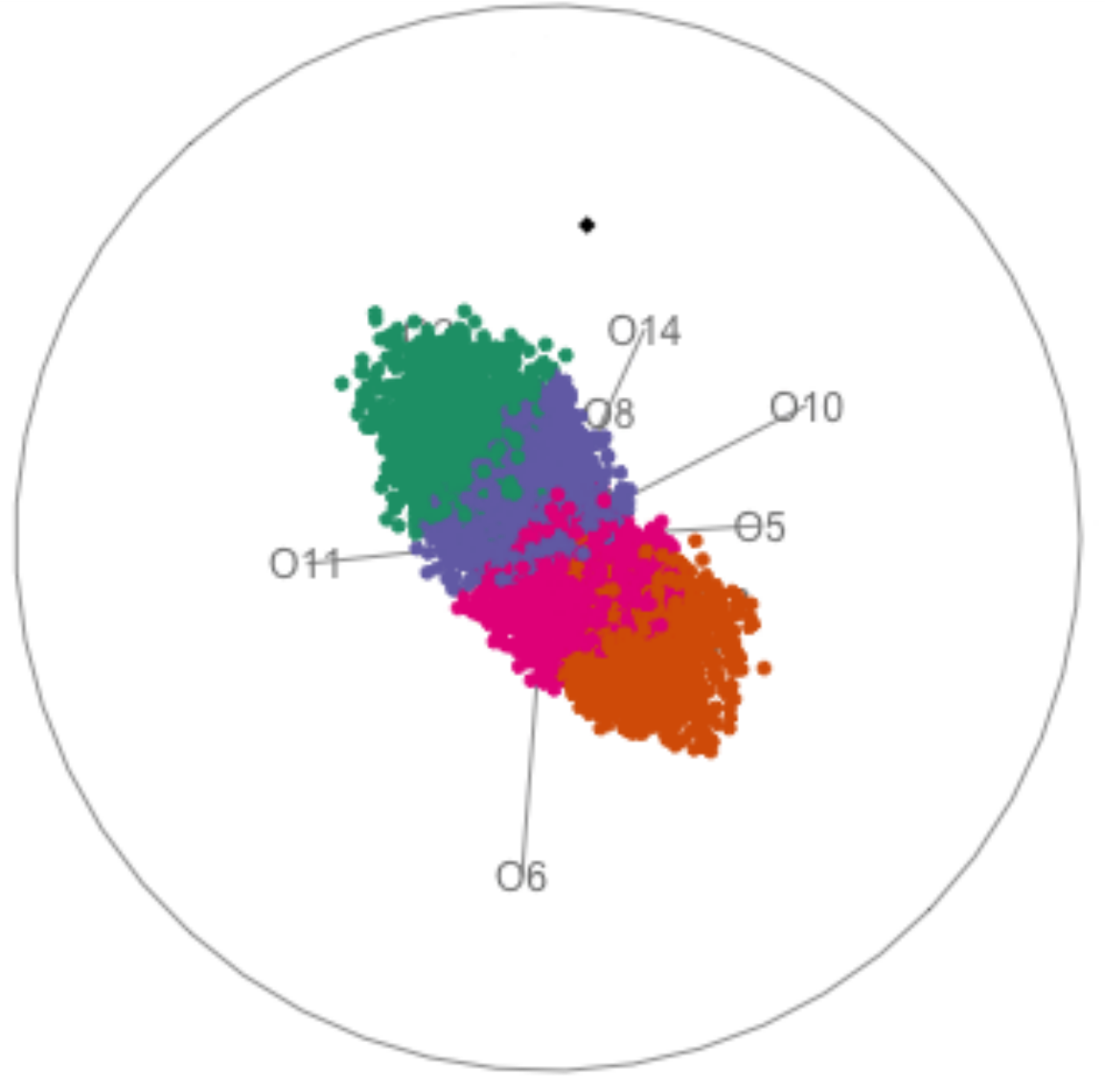}}
\caption{Selected projections from tours in parameter space (left and center) and observable space (right) of the clusters resulting with four parameters. The color code matches the one in the right panel of Fig.~\ref{f:2dvs4d}.}
\label{f:4dcase}
\end{figure}

We now turn to visualize the parameter space for this 4D case. In Fig.~\ref{f:4dcase} we show on the left panel a $C_9-C_{10}$ projection which shows how the correlations between $C_9-C_{10}$ due to $R_K$ are still dominant. The right panel shows a projection from observable space where it is clear that this 4D volume of models also does not contain the experimental point. The center panel is a thin slice projected onto the $C_9-C_{9^\prime}$ that illustrates correlations between these two parameters that are not visible without slicing, obtained with the tool described in \cite{mmtour}. The clusters in parameter and observable spaces for this case can be better visualized with animations~3~and~4. Animation~5 shows the effect of slicing through the SM point and projecting onto the $C_9-C_{10}$ plane (the interactive tool {\tt mmtour} allows one to change the slice height and the projection plane). Animation~6 shows what happens when varying the slice height while projecting onto the $C_9-C_{9^\prime}$ plane. The latter reveals correlations between these two parameters that are only visible in thin slices and obscured in any projection.

\section{Including more observables}

\begin{figure}[h]
\centering{
\includegraphics[scale=0.6]{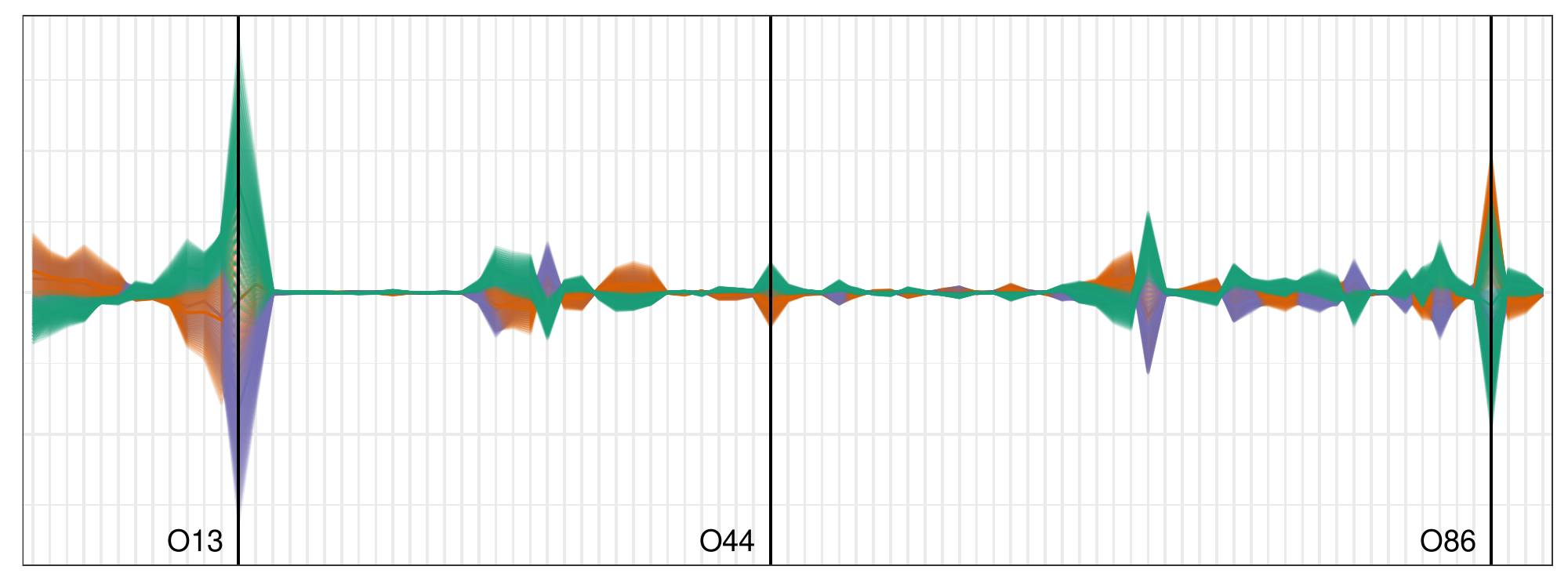}}
\caption{Centered PC plot for the 89 observables listed in \cite{Laa:2021dlg}, the first 14 correspond to those in Table~\ref{t:obs}.}
\label{f:cenpc89}
\end{figure}

\begin{figure}[h]
\centering{
\includegraphics[scale=0.32]{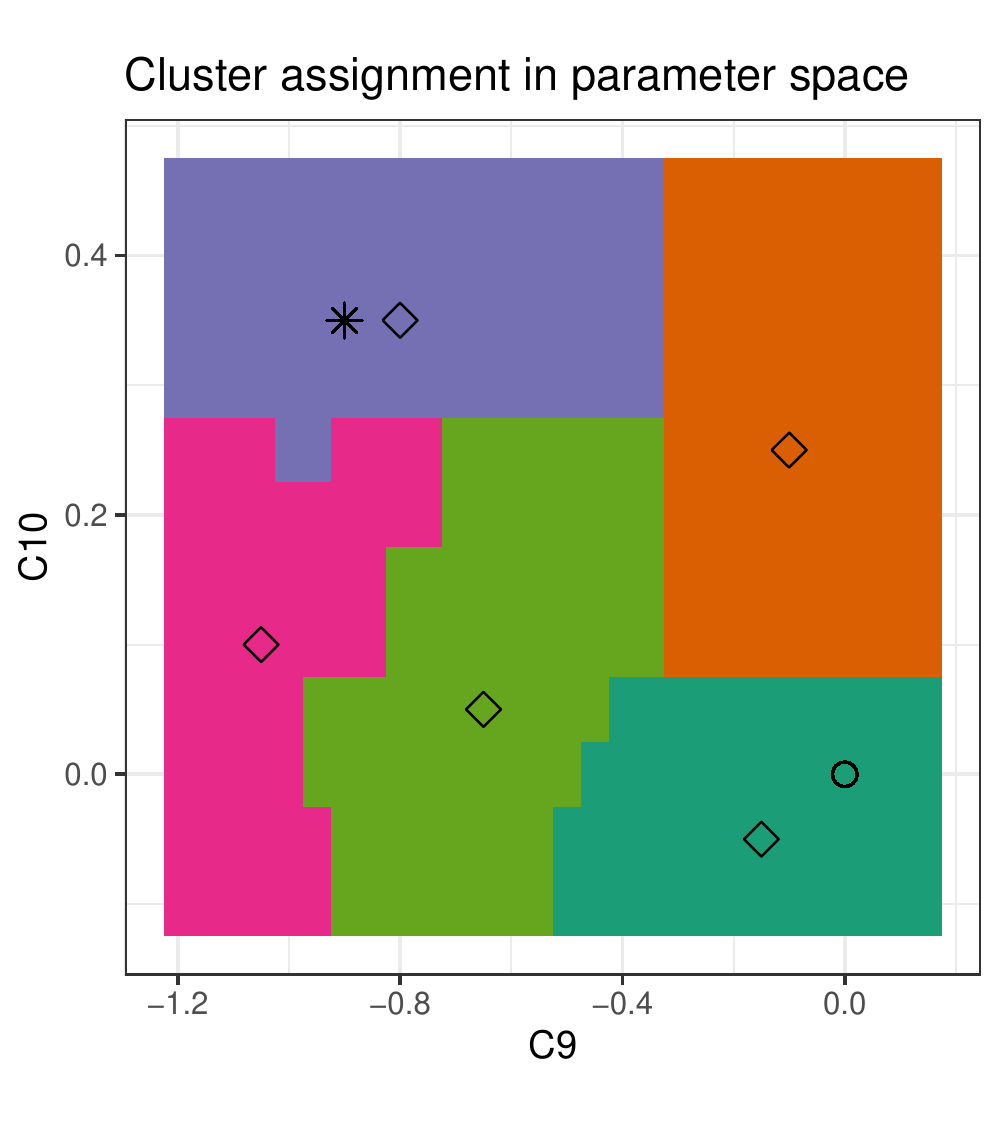}\includegraphics[scale=0.32]{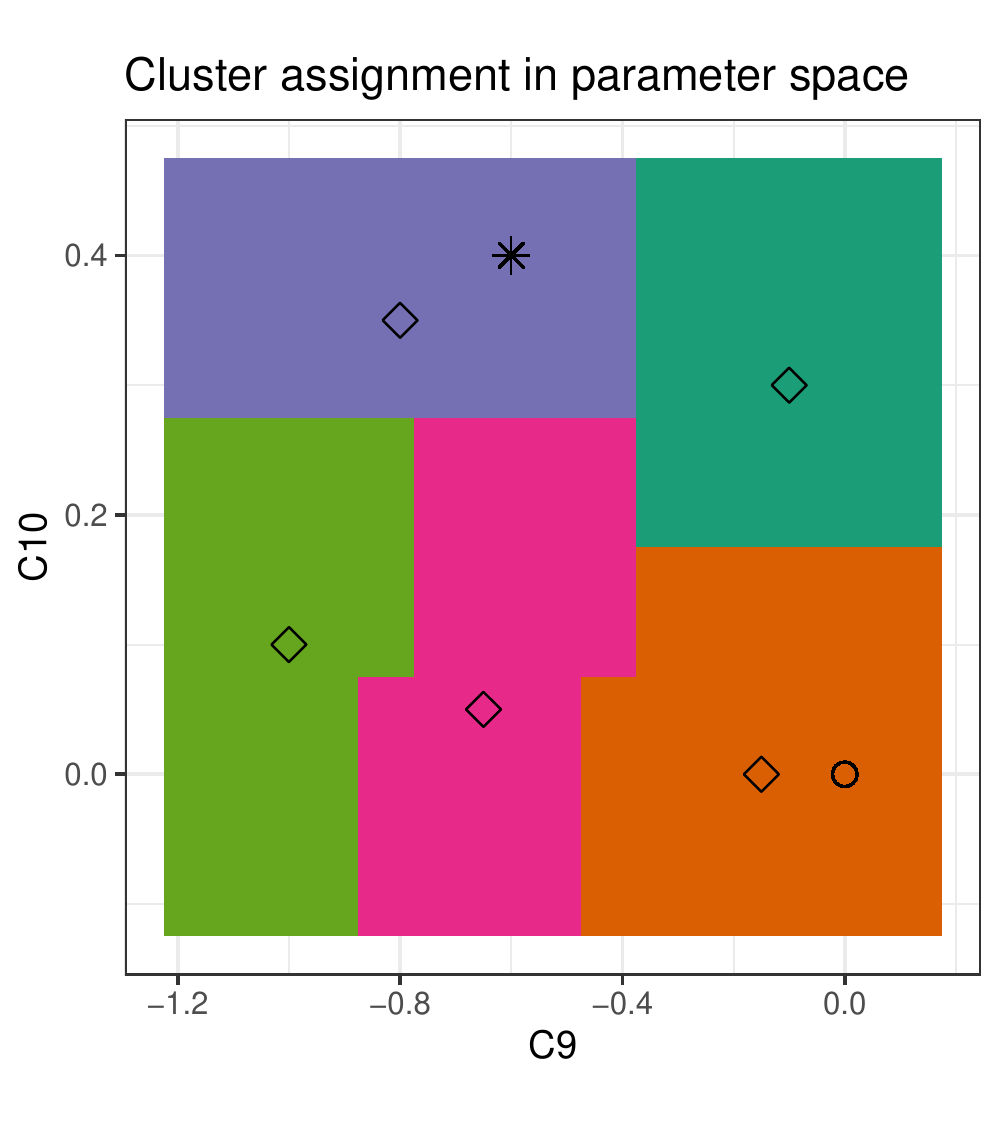}\includegraphics[scale=0.32]{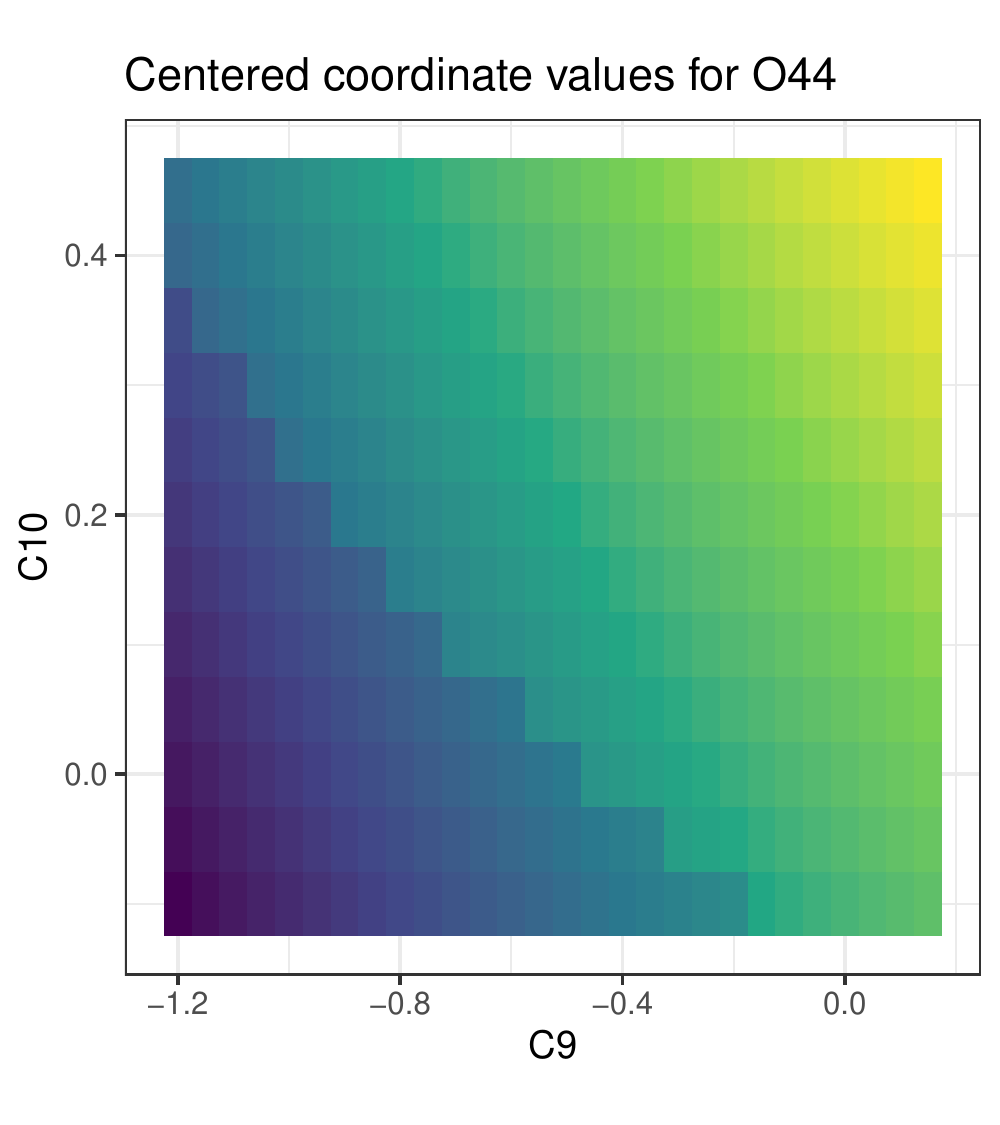}\includegraphics[scale=0.32]{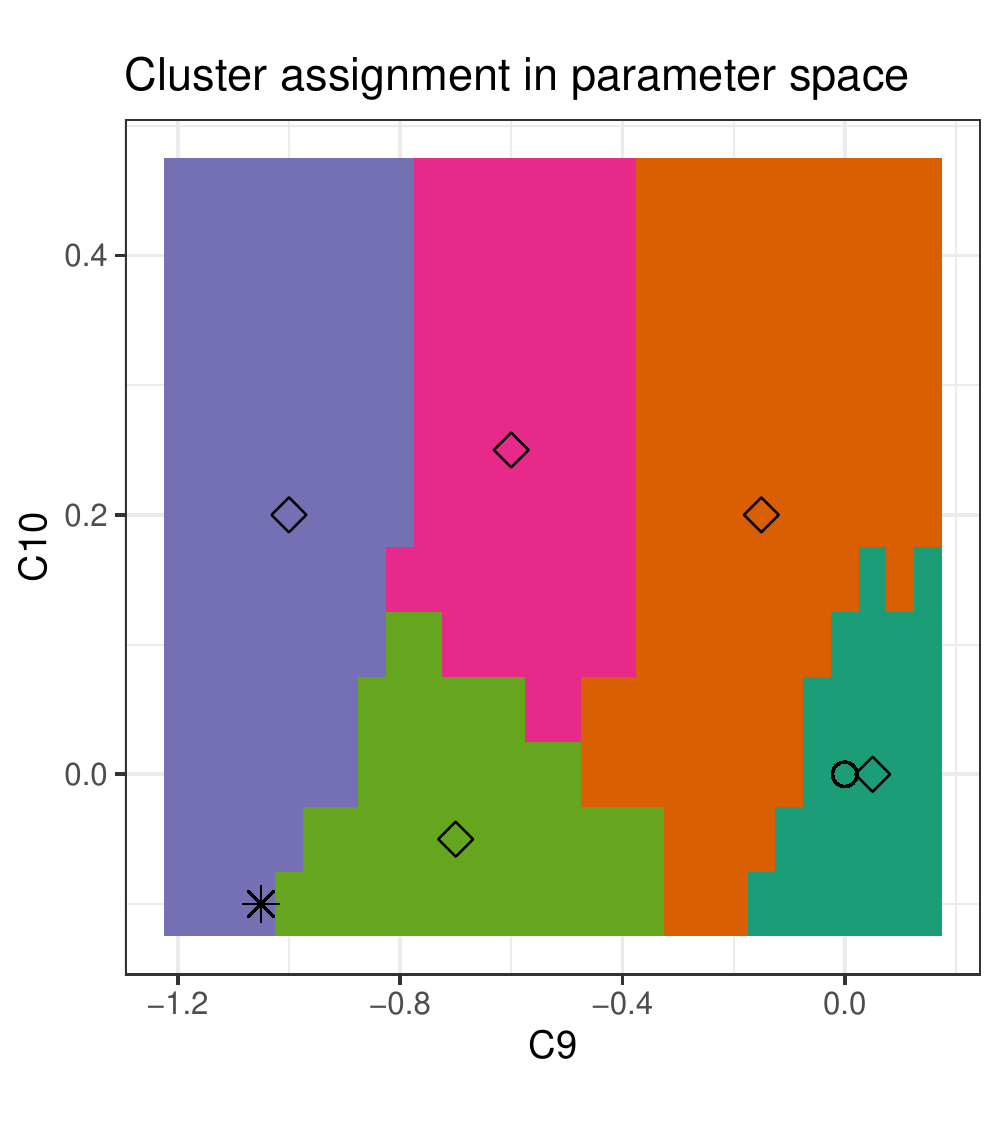}}
\caption{The left panel shows 5 clusters in parameter space with only two parameters, $C_9,C_{10}$, and 89 observables. The second pane from the left shows the 5 clusters including only the first 14 observables plus $B(B_s\to\mu^+\mu^-)$ as described in the text. The third panel shows the variation of $O_{44}$ with $C_9,C_{10}$ and the last panel the 5 clusters that would be obtained using only the first 14 observables plus $P_4^\prime[0.1-0.98]$ with an experimental error four times smaller than it currently is.}
\label{f:wc8986}
\end{figure}

As we know hundreds of observables have been discussed in connection with the $b\to s\ell^+\ell^-$ transitions. Here we look at the 89 that we selected in \cite{Laa:2021dlg}, with the first 14 being those in Table~\ref{t:obs}. Using all of them, the resolving power of this data set is between 8 and 10 clusters. We will illustrate the main results using only five clusters. The centered PC plot of Fig.~\ref{f:cenpc89} can be used to select additional ones that may be important. In particular $O_{86}$ ($B(B_s\to\mu^+\mu^-)$) stands out.  
If we use the average experimental error computed by {\tt flavio}, $B(B_s\to\mu^+\mu^-)=(2.81\pm0.24)\times 10^{-9}$, this observable alone explains most of the difference in the clusters obtained with the set of 89 observables and with only the first 14. This can be seen by comparing the left two panels in Fig.~\ref{f:wc8986}. 

For a different application of these results, we turn our attention to $O_{44}$ which Fig.~\ref{f:cenpc89} shows to have moderate importance. In the third panel of Fig.~\ref{f:wc8986} we show the coordinate variation of this observable. It suggests that it can constrain directions missed by the current overall picture if its significance can be enhanced. Currently, this observable has the experimental value $O_{44}=P_4^\prime[0.1-0.98]=0.135\pm0.118$. We can study what happens if the uncertainty in this measurement can be reduced in the future. For example, the right panel of Fig.~\ref{f:wc8986} shows the effect of adding just this observable to the original set of 14 but assumes that its  experimental error can be {\bf reduced by a factor of four}. 

\section{Conclusions}

Using the example of the B anomalies we have demonstrated how to investigate the relationship between parameter and observable space using a group of related displays to interpret different clustering outcomes. This analysis is facilitated by the interactive environment of the tool {\tt pandemonium}, which allows for easy comparison of clustering results with different parameter settings. By choosing different settings, specific observables can be emphasized or suppressed. The tool provides information to decide what is the optimal number of partitions for a given data set, and which observables should be emphasized to explore specific directions in parameter space. In this talk, we applied these methods to discuss a well-known B physics problem, which provides feedback for using these methods for other cases.

For the B anomalies example, our study highlights the importance of $R_K$ and how this is connected to the precision of the measurement. With the new, more precise, measurement this observable becomes even more dominant. Even though global fits will be closer to the SM with the new values of $R_K$ and $R_{K^\star}$, the tension with certain bins of $P_5^\prime$ is increased. Increased precision in different observables is required to explore directions of parameter space that are orthogonal to $R_K$, and our tool helps identify which ones to emphasize for different purposes. The observable space tours reveal when the parameters of a model are insufficient to address a discrepancy with a reference point and PC plots can be used to quantify this. In this example, using four WC improves but does not solve the inability of models parametrized in this form to completely address the experimental results. We introduced tours with slicing to allow for the examination of correlations between parameters that are hidden in projections, in particular between $C_9$ and $C_{9^\prime}$. This situation occurs when the correlation is obscured by the effect of other parameters. 
 
 The technique and software tools we propose have broad applications in particle physics and other fields, offering new insights, particularly in less understood scenarios. 
 
 The tools described in this talk have been implemented by Ursula Laa and are available on GitHub as a Shiny app (an R package) at https://github. com/uschiLaa/pandemonium. The tools used for the slice tour shown in this talk have been implemented by Alex Aumann as a Mathematica package available on GitHub at https://github.com/uschiLaa/mmtour.

\acknowledgments{This work was supported in part by the Australian Government through the Australian Research Council.}
\bibliography{biblio}

\providecommand{\href}[2]{#2}\begingroup\raggedright\begin{thebibliography}{10}

\bibitem{Laa:2021dlg}
U.~Laa and G.~Valencia, \emph{{Pandemonium: a clustering tool to partition
  parameter space\textemdash{}application to the B anomalies}},
  \href{http://dx.doi.org/10.1140/epjp/s13360-021-02310-1}{\emph{Eur. Phys. J.
  Plus} {\bfseries 137} (2022) 145},
  [\href{https://arxiv.org/abs/2103.07937}{{\ttfamily 2103.07937}}].

\bibitem{LHCb:2022qnv}
{\scshape LHCb} collaboration, \emph{{Test of lepton universality in $b
  \rightarrow s \ell^+ \ell^-$ decays}},
  \href{https://arxiv.org/abs/2212.09152}{{\ttfamily 2212.09152}}.

\bibitem{Capdevila:2018jhy}
B.~Capdevila, U.~Laa and G.~Valencia, \emph{{Anatomy of a six-parameter fit to
  the $b\to s \ell^+\ell^-$ anomalies}},
  \href{http://dx.doi.org/10.1140/epjc/s10052-019-6944-8}{\emph{Eur. Phys. J.
  C} {\bfseries 79} (2019) 462},
  [\href{https://arxiv.org/abs/1811.10793}{{\ttfamily 1811.10793}}].

\bibitem{Straub:2018kue}
D.~M. Straub, \emph{{flavio: a Python package for flavour and precision
  phenomenology in the Standard Model and beyond}},
  \href{https://arxiv.org/abs/1810.08132}{{\ttfamily 1810.08132}}.

\bibitem{Aaij:2020nrf}
{\scshape LHCb} collaboration, R.~Aaij et~al., \emph{{Measurement of
  $CP$-Averaged Observables in the $B^{0}\rightarrow K^{*0}\mu^{+}\mu^{-}$
  Decay}}, \href{http://dx.doi.org/10.1103/PhysRevLett.125.011802}{\emph{Phys.
  Rev. Lett.} {\bfseries 125} (2020) 011802},
  [\href{https://arxiv.org/abs/2003.04831}{{\ttfamily 2003.04831}}].

\bibitem{Sirunyan:2017dhj}
{\scshape CMS} collaboration, A.~M. Sirunyan et~al., \emph{{Measurement of
  angular parameters from the decay $\mathrm{B}^0 \to \mathrm{K}^{*0} \mu^+
  \mu^-$ in proton-proton collisions at $\sqrt{s} = $ 8 TeV}},
  \href{http://dx.doi.org/10.1016/j.physletb.2018.04.030}{\emph{Phys. Lett. B}
  {\bfseries 781} (2018) 517--541},
  [\href{https://arxiv.org/abs/1710.02846}{{\ttfamily 1710.02846}}].

\bibitem{Aaboud:2018krd}
{\scshape ATLAS} collaboration, M.~Aaboud et~al., \emph{{Angular analysis of
  $B^0_d \rightarrow K^{*}\mu^+\mu^-$ decays in $pp$ collisions at $\sqrt{s}=
  8$ TeV with the ATLAS detector}},
  \href{http://dx.doi.org/10.1007/JHEP10(2018)047}{\emph{JHEP} {\bfseries 10}
  (2018) 047}, [\href{https://arxiv.org/abs/1805.04000}{{\ttfamily
  1805.04000}}].

\bibitem{Aaij:2019wad}
{\scshape LHCb} collaboration, R.~Aaij et~al., \emph{{Search for
  lepton-universality violation in $B^+\to K^+\ell^+\ell^-$ decays}},
  \href{http://dx.doi.org/10.1103/PhysRevLett.122.191801}{\emph{Phys. Rev.
  Lett.} {\bfseries 122} (2019) 191801},
  [\href{https://arxiv.org/abs/1903.09252}{{\ttfamily 1903.09252}}].

\bibitem{Abdesselam:2019lab}
{\scshape Belle} collaboration, A.~Abdesselam et~al., \emph{{Test of lepton
  flavor universality in $B \to K \ell^{+}\ell^{-}$ decays}},
  \href{https://arxiv.org/abs/1908.01848}{{\ttfamily 1908.01848}}.

\bibitem{Aaij:2017vbb}
{\scshape LHCb} collaboration, R.~Aaij et~al., \emph{{Test of lepton
  universality with $B^{0} \rightarrow K^{*0}\ell^{+}\ell^{-}$ decays}},
  \href{http://dx.doi.org/10.1007/JHEP08(2017)055}{\emph{JHEP} {\bfseries 08}
  (2017) 055}, [\href{https://arxiv.org/abs/1705.05802}{{\ttfamily
  1705.05802}}].

\bibitem{Abdesselam:2019wac}
{\scshape Belle} collaboration, A.~Abdesselam et~al., \emph{{Test of lepton
  flavor universality in ${B\to K^\ast\ell^+\ell^-}$ decays at Belle}},
  \href{https://arxiv.org/abs/1904.02440}{{\ttfamily 1904.02440}}.

\bibitem{Laa:2019bap}
U.~Laa, D.~Cook and G.~Valencia, \emph{A slice tour for finding hollowness in
  high-dimensional data},
  \href{http://dx.doi.org/10.1080/10618600.2020.1777140}{\emph{Journal of
  Computational and Graphical Statistics} {\bfseries 29} (2020) 681--687},
  [\href{https://arxiv.org/abs/https://doi.org/10.1080/10618600.2020.1777140}{{\ttfamily
  https://doi.org/10.1080/10618600.2020.1777140}}].

\bibitem{Laa:2020wkm}
U.~Laa, D.~Cook, A.~Buja and G.~Valencia, \emph{Hole or grain? a section
  pursuit index for finding hidden structure in multiple dimensions},
  \href{http://dx.doi.org/10.1080/10618600.2022.2035230}{\emph{Journal of
  Computational and Graphical Statistics} {\bfseries 31} (2022) 739--752},
  [\href{https://arxiv.org/abs/https://doi.org/10.1080/10618600.2022.2035230}{{\ttfamily
  https://doi.org/10.1080/10618600.2022.2035230}}].

\bibitem{mmtour}
U.~Laa, A.~Aumann, D.~Cook and G.~Valencia, \emph{{New and simplified manual
  controls for projection and slice tours, with application to exploring
  classification boundaries in high dimensions}},
  \href{https://arxiv.org/abs/2210.05228}{{\ttfamily 2210.05228}}.

\end{thebibliography}\endgroup
\end{document}